\documentclass{aa}  

\usepackage{graphicx}
\usepackage{txfonts}
\usepackage{longtable}
\usepackage{rotating}
\usepackage{natbib}
\usepackage{graphics}
\usepackage{psfrag}
\usepackage{amssymb}
\usepackage{xspace}
\usepackage{color}
\usepackage{mathtools}
\bibpunct{(}{)}{;}{a}{}{,}
\def\ha{H$\alpha$}
\def\hb{H$\beta$}
\begin{document}

\title{A data-driven approach to constraining the atmospheric temperature structure of the ultra-hot Jupiter KELT-9b}
\subtitle{}

\titlerunning{A data-driven approach to constraining the atmospheric temperature structure of KELT-9b}
\authorrunning{Fossati et al.}

\author{L. Fossati\inst{1}	\and
	D. Shulyak\inst{2}	\and
	A. G. Sreejith\inst{1}	\and
	T. Koskinen\inst{3}	\and
	M. E. Young\inst{1}	\and
	P. E. Cubillos\inst{1}	\and
	L. M. Lara\inst{4}	\and
	K. France\inst{5,6}	\and
	M. Rengel\inst{2}	\and
	P. W. Cauley\inst{5}	\and
	J. D. Turner\inst{7}	\and
	A. Wyttenbach\inst{8}	\and
	F. Yan\inst{9}
}
\institute{
Space Research Institute, Austrian Academy of Sciences, Schmiedlstrasse 6, A-8042 Graz, Austria\\
\email{luca.fossati@oeaw.ac.at}
\and 
Max-Planck Institut f\"ur Sonnensystemforschung, Justus-von-Liebig-Weg 3, D-37077, G\"ottingen, Germany
\and
Lunar and Planetary Laboratory, University of Arizona, 1629 East University Boulevard, Tucson, AZ 85721-0092, USA
\and
Instituto de Astrof\'isica de Andaluc\'ia - CSIC, c/ Glorieta de la Astronom\'ia s/n, 18008 Granada, Spain
\and
Laboratory for Atmospheric and Space Physics, University of Colorado, 600 UCB, Boulder, CO 80309, USA
\and
Center for Astrophysics and Space Astronomy, University of Colorado, 389 UCB, Boulder, CO 80309, USA
\and
Department of Astronomy, Cornell University, Ithaca NY USA
\and
Universit\'e Grenoble Alpes, CNRS, IPAG, 38000 Grenoble, France
\and
Institut f\"ur Astrophysik, Georg-August-Universit\"at, Friedrich-Hund-Platz 1, D-37077 G\"ottingen, Germany
}

\date{}

 
\abstract
{Observationally constraining the atmospheric temperature-pressure (TP) profile of exoplanets is an important step forward for improving planetary atmosphere models, further enabling one to place the detection of spectral features and the measurement of atomic and molecular abundances through transmission and emission spectroscopy on solid ground.}
{The aim is to constrain the TP profile of the ultra-hot Jupiter KELT-9b by fitting synthetic spectra to the observed \ha\ and \hb\ lines and identify why self-consistent planetary TP models are unable to fit the observations.}
{We construct 126 one-dimensional TP profiles varying the lower and upper atmospheric temperatures, as well as the location and gradient of the temperature rise. For each TP profile, we compute transmission spectra of the \ha\ and \hb\ lines employing the Cloudy radiative transfer code, which self-consistently accounts for non-local thermodynamic equilibrium (NLTE) effects.}
{The TP profiles leading to best fit the observations are characterised by an upper atmospheric temperature of 10000--11000\,K and by an inverted temperature profile at pressures higher than 10$^{-4}$\,bar. We find that the assumption of local thermodynamic equilibrium (LTE) leads to overestimate the level population of excited hydrogen by several orders of magnitude, and hence to significantly overestimate the strength of the Balmer lines. The chemical composition of the best fitting models indicate that the high upper atmospheric temperature is most likely driven by metal photoionisation and that Fe{\sc ii} and Fe{\sc iii} have comparable abundances at pressures lower than 10$^{-6}$\,bar, possibly making the latter detectable.}
{Modelling the atmospheres of ultra-hot Jupiters requires one to account for metal photoionisation. The high atmospheric mass-loss rate ($>$10$^{11}$\,g\,s$^{-1}$), caused by the high temperature, may have consequences on the planetary atmospheric evolution. Other ultra-hot Jupiters orbiting early-type stars may be characterised by similarly high upper atmospheric temperatures and hence high mass-loss rates. This may have consequences on the basic properties of the observed planets orbiting hot stars.}

\maketitle
%
%
\section{Introduction} \label{sec:introduction}
More than twenty years of exoplanet research has greatly broadened and deepened our understanding of planets. In particular, the detection of close-in giant planets has opened the possibility of characterising planetary atmospheres \citep[e.g.,][]{seager2000,brown2001,charbonneau2002,snellen2010,brogi2019}. Atmospheric characterisation has become one of the central aspects of exoplanet science such that current (e.g., TESS) and future (e.g., PLATO) planet detection facilities aim at finding planets orbiting stars bright enough to enable transmission and/or emission spectroscopy \citep[e.g.,][]{rauer2014,ricker2015}.

The atmospheric characterisation observations conducted so far have led to the detection of a wide range of atomic and molecular species in exoplanetary atmospheres \citep[e.g.,][]{vidal-madjar2004,redfield2008,fossati2010,deming2013,barman2015,kreidberg2015,line2016,sing2016,casasays-barris2017,evans2017,lendl2017,line2017,birkby2017,wyttenbach2017,brogi2018,jensen2018,nikolov2018,nortmann2018,tsiaras2018,wakeford2018,hoeijmakers2019,vonessen2019}. Furthermore, the community has devised sophisticated forward model and retrieval techniques to extract relevant physical information (e.g., temperature structure, abundance profiles) from the observations \citep[e.g.,][]{irwin2008,madhu2009,madhu2010,howe2012,lee2012,line2013,waldmann2015a,waldmann2015b,lavie2017,macdonald2017,malik2017,molliere2015,molliere2019,shulyak2019,barstow2013,barstow2017,barstow2020}. The main general result is that exoplanetary atmospheres present a large variety of physical and chemical properties. The picture is further complicated by the fact that most planets host aerosols hiding spectral features and/or sequestering elements producing detectable features \citep[e.g.,][]{sing2016,parmentier2016,gibson2017,wakeford2017,benneke2019}. Therefore, for some hot Jupiters, a thorough observational atmospheric characterisation may be limited to the upper atmospheric layers.

However, ultra-hot Jupiters, gas giant exoplanets with equilibrium temperatures exceeding 2000\,K and typically orbiting intermediate-mass stars (A- and early F-type), are not subject to this limitation, because their high atmospheric temperature and intense irradiation ensure that the molecules composing aerosols are dissociated, at least on the day-side \citep[e.g.,][]{parmentier2018,kitzmann2018,lothringer2018}. About a dozen ultra-hot Jupiters have been detected to date and the most studied one is KELT-9b, also known as HD\,195689\,b, which is the hottest planet orbiting a non-degenerate star among those found so far \citep{gaudi2017}. Both day- and night-side temperatures of KELT-9b have been measured through phase curve observations at optical (with TESS) and infrared (with Spitzer at 4.5\,$\mu$m) wavelengths. From TESS photometry, \citet{wong2019} obtained a night-side temperature of 3020$\pm$90\,K and a day-side temperature of 4570$\pm$90\,K, while from Spitzer data \citet{mansfield2020} derived a night-side temperature of 2556$^{+101}_{-97}$\,K and a day-side temperature of 4566$^{+140}_{-136}$\,K. Both hydrogen and several metals have been detected in the atmosphere of KELT-9b through ground-based high-resolution transmission spectroscopy \citep[e.g.,][]{yan2018,yan2019,hoeijmakers2018,borsa2019,hoeijmakers2019,cauley2019,turner2020,pino2020,wyttenbach2020}. The detection of metals, in particular of ions such as Fe{\sc ii}, indicates that the atmospheric temperature in the region where lines of these elements form is above 4000\,K \citep{hoeijmakers2019}.

Self-consistent atmospheric modelling of \object{KELT-9b} conducted with the PHOENIX stellar and planetary model atmosphere code indicates that the temperature profile should be inverted, reaching temperatures of the order of 6300\,K in the upper atmosphere \citep{lothringer2018,fossati2018b}. \citet{pino2020} detected such a thermal inversion from secondary eclipse high-resolution spectroscopic observations. \citet{garcia2019} modelled the upper atmosphere of KELT-9b assuming a pure hydrogen composition accounting for stellar irradiation, with a particular focus on the heating caused by the near-ultraviolet photons, and accounting for non-local thermodynamic equilibrium (NLTE) effects. They obtained a temperature profile reaching a maximum of about 15000\,K at a pressure of 10$^{-9}$\,bar, but mentioned that the implementation of cooling by Fe{\sc ii} would decrease the temperature by 1000--2000\,K, presumably more with the inclusion of additional metals. They also showed that in their model NLTE effects play a significant role in shaping the properties of the planetary upper atmosphere.

The \ha\ hydrogen Balmer line has been the first one detected in the atmosphere of KELT-9b, shortly followed by some of the higher-order hydrogen Balmer lines \citep[e.g.,][]{yan2018,cauley2019,turner2020,wyttenbach2020}. The detection of the \ha\ line revealed that the planet hosts an extended, hot hydrogen envelope, while the large transit depth gave rise to the idea that \ha\ could be used to directly probe the extended, escaping planetary upper atmosphere, hence constrain mass loss \citep{yan2018,fossati2018b,garcia2019}. \citet{turner2020} showed that the \ha\ line does not probe the planetary atmosphere beyond the Roche lobe, however this does not prevent constraining the atmospheric mass-loss rate \citep{yan2018,fossati2018b,wyttenbach2020}. As a matter of fact, the detection and modelling of the hydrogen Balmer lines enable one to constrain the atmospheric temperature between about the mbar and the nbar pressure level \citep{garcia2019,wyttenbach2020}, hence the energetics driving the escape.

\citet{garcia2019}, \citet{turner2020}, and \citet{wyttenbach2020} employed the observed \ha\ line profile of KELT-9b to extract information on the planetary atmospheric properties. \citet{garcia2019} and \citet{turner2020} showed that the \ha\ synthetic transmission spectrum computed accounting for NLTE effects and with the system parameters obtained by \citet{gaudi2017} was significantly weaker than the observed one. \citet{garcia2019} concluded that this was possibly the result of an overestimation of the planetary mass, particularly because at that time the mass measurement was affected by a large uncertainty \citep[2.88$\pm$0.84\,$M_{\rm Jup}$;][]{gaudi2017}. However, additional and more accurate radial velocity measurements confirmed the planetary mass derived by \citet{gaudi2017}, further providing a significantly smaller uncertainty \citep[2.88$\pm$0.35\,$M_{\rm Jup}$;][]{borsa2019}. In order to fit the observed \ha\ line profile, \citet{turner2020} employed a smaller planetary mass by about 30\%, but they suggested that the line profile may be also reproduced by increasing the temperature compared to that provided by PHOENIX, instead of decreasing the planetary mass. \citet{wyttenbach2020} employed a retrieval-like technique to reconstruct the atmospheric temperature, mass-loss rate, and density of excited hydrogen by fitting the observed \ha, \hb, and H$\gamma$ lines. They assumed an isothermal profile and a constant density ratio of excited hydrogen to total hydrogen, and did not consider photoionisation. By keeping the hydrogen density profile equal to that given by the Boltzmann equation, hence assuming local thermodynamic equilibrium (LTE), they obtained an atmospheric temperature of 13200$^{+800}_{-720}$\,K. Instead, by leaving the density of excited hydrogen as a free parameter, they obtained an atmospheric temperature of 9600$\pm$1200\,K and a density of excited hydrogen relative to the total amount of hydrogen of about 10$^{-11}$. However, there are significant degeneracies among the free parameters.

In this work, we construct a large number of atmospheric temperature-pressure (TP) profiles, ranging from about 1 to 10$^{-11}$\,bar, and employ the Cloudy NLTE radiative transfer code \citep{ferland2017} to constrain the TP profile, or family of TP profiles, that best reproduce the available transmission spectroscopy observations of the \ha\ and \hb\ lines. Observationally constraining the planetary atmospheric TP profile is an important step towards the characterisation of the prototype ultra-hot Jupiter and it significantly advances models of the planetary upper atmosphere, further helping to place ultra-hot Jupiter escape on more solid observational ground. The knowledge, even if approximate, of the atmospheric TP profile would also enable one to produce more accurate synthetic transmission spectra to be employed for detecting metals in the planetary atmosphere, for example, to refine the inference of the local thermodynamic state and to better constrain physico-chemical processes in the atmospheres of ultra-hot Jupiters.

This paper is organised as follows. In Sect.~\ref{sec:data}, we present the considered observational material and compare the available \ha\ and \hb\ line profiles with each other. Section~\ref{sec:tp} describes how we computed the synthetic TP profiles, while Sect.~\ref{sec:transmission} presents the modelling scheme we employed to calculate the synthetic transmission spectra from the TP profiles. Section~\ref{sec:results} presents the results that we discuss in Sect.~\ref{sec:discussion}. We gather the conclusions of this work in Sect.~\ref{sec:conclusion}.
\section{Observed transmission spectra}\label{sec:data}
Transmission spectra of KELT-9b covering at least one of the hydrogen Balmer lines, typically \ha, have been published by \citet{yan2018}, \citet{cauley2019}, \citet{turner2020}, and \citet{wyttenbach2020}. Both \citet{yan2018} and \citet{turner2020} obtained the \ha\ transmission spectrum using the CARMENES high-resolution \'echelle spectrograph ($R$\,$\approx$\,95,000) attached to the 3.5\,m telescope of the Calar Alto Observatory. The transmission spectrum of \citet{yan2018} was obtained combining transits observed during two nights in 2017 (August and September), while that of \citet{turner2020} was obtained from one transit observation recorded in 2018 (June). \citet{cauley2019} obtained the \ha\ and \hb\ transmission spectra of KELT-9b following one transit observation collected in 2018 (July) using the PEPSI high-resolution \'echelle spectrograph ($R$\,$\approx$\,50,000) attached to the Large Binocular Telescope. We employ here the transmission spectrum obtained considering only the first third of the transit observed by \citet{cauley2019}, because the rest of the transit appears to be characterised by variability. \citet{wyttenbach2020} obtained transmission spectra of the \ha\ to H$\delta$ lines following two transit observations conducted in July 2017 and July 2018 with the HARPS-N high-resolution spectrograph ($R$\,$\approx$\,115,000) attached to the 3.6\,m Telescopio Nazionale Galileo. 

\citet{yan2018} and \citet{cauley2019} normalised the already blaze-corrected spectra using linear fits to adjacent continuum points, finally stitching the spectral orders together. They further corrected telluric contamination employing a theoretical telluric spectrum, finally dividing each spectrum by the average out-of-transit spectrum and then shifting all spectra to the planet rest frame before co-adding the in-transit residual spectra. \citet{turner2020} applied SYSREM \citep{tamuz2005} to remove telluric and stellar lines and other systematics from the blaze-corrected spectra, which have then shifted to the planet rest frame before co-adding. \citet{wyttenbach2020} followed spectral reduction and analysis procedures similar to those of \citet{yan2018} and \citet{cauley2019}, but they extracted the transmission spectrum computing the difference between in- and out-of-transit spectra to correct for the Rossiter-McLaughlin (RM) effect, and only in a second step, the residuals have been divided by the average out-of-transit spectrum. All authors corrected the data for the RM effect, while \citet{yan2018} and \citet{cauley2019} further corrected the data for the Center-to-Limb Variation (CLV) effect \citep[see, for example][]{yan2017}.

\begin{figure}[h!]
\vspace{0.5cm}
\includegraphics[width=\hsize,clip]{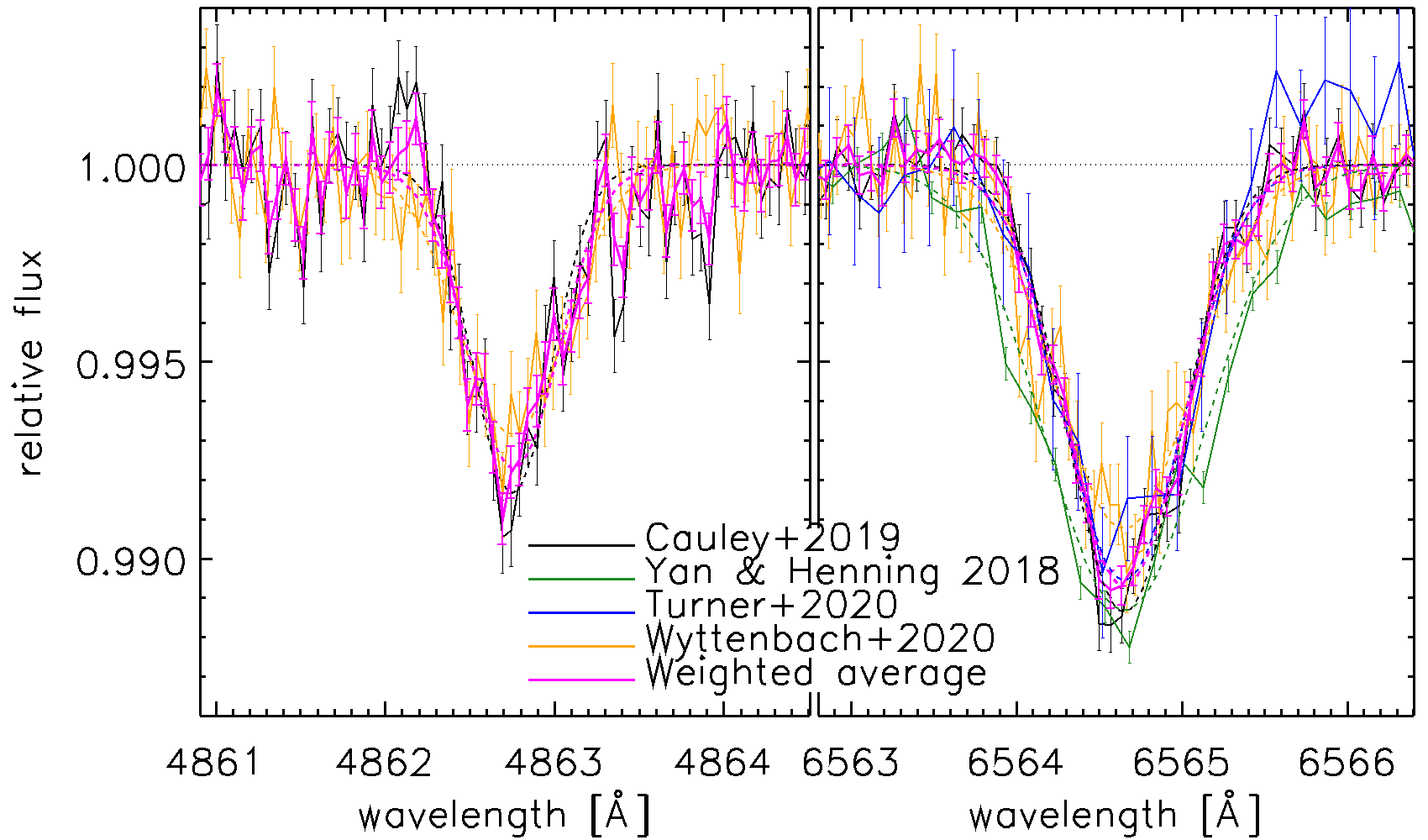}
\caption{Transmission spectra of the hydrogen Balmer line profiles (\hb, left; \ha, right) published by \citet[][green]{yan2018}, \citet[][black]{cauley2019}, \citet[][blue]{turner2020}, and \citet[][orange]{wyttenbach2020}. The magenta line is the weighted average profile obtained considering just the profiles of \citet{cauley2019} and \citet{wyttenbach2020}. Wavelengths are in vacuum to match those of the synthetic transmission spectra (see Sect.~\ref{sec:transmission}). The data have been rebinned by a factor of five for visualisation purposes. The dashed lines are Gaussian fits to the data. The horizontal dotted line at one sets the continuum level to guide the eye.}
\label{fig:comparison_observations}
\end{figure}

Figure~\ref{fig:comparison_observations} compares the \ha\  (right panel) and \hb\  (left panel) transmission spectra published by \citet{yan2018}, \citet{cauley2019}, \citet{turner2020}, and \citet{wyttenbach2020}. To ease the comparison, in the making of Fig.~\ref{fig:comparison_observations} we removed the small wavelength shifts present among the different line profiles, which are probably caused by the use of different systemic radial velocity values. Because of differences in the number of observed transits or in the collecting area of the telescopes employed to obtain the data, the noise level in the transmission spectra of \citet{yan2018}, \citet{cauley2019}, and \citet{wyttenbach2020}, as indicated by the error bars, is significantly smaller than that of \citet{turner2020}.

The measurements of the full width at half maximum (FWHM) given by the Gaussian fits indicate that there are significant differences among the observed line widths \citep[see also][for a similar discussion]{wyttenbach2020}. As indicated in Table~\ref{tab:observations}, the FWHM obtained from the \ha\ profile of \citet{yan2018} is about 1.5 times larger than those obtained from the profiles of \citet{cauley2019} and \citet{turner2020}, while that of \citet{wyttenbach2020} lies in between those. Also for the H$\beta$ line, the FWHM derived from the profile of \citet{wyttenbach2020} is significantly larger than that obtained from the profile of \citet{cauley2019}. The \ha\ and \hb\ profiles of \citet{wyttenbach2020} are significantly shallower than those of the other authors. These differences are not due to differences in the spectral resolution of the instruments/data, but may stem out of differences in the data reduction and analysis procedures, although for example \citet{cauley2019} accounts for the CLV effect, while \citet{turner2020} does not, but the two profiles are comparable. Differences may, however, rise also from the treatment of the Doppler shadow \citep{borsa2019,wyttenbach2020} or from the fact that the different authors considered significantly different systemic velocities. Such differences could also rise as a result of instrumental systematics different for each instrument and that have not been properly corrected for. Finally, it is also possible that the observed differences are of astrophysical nature and due to intrinsic variations of the hydrodynamically escaping atmosphere. Indeed, \citet{cauley2019} reported on the possible presence of transit depth variations along a single transit. Table~\ref{tab:observations} also shows that the equivalent widths derived from the profiles of \citet{wyttenbach2020} and \citet{cauley2019} are comparable, which is because the former are shallower and broader, while the latter are deeper and narrower. 

For the comparison of the synthetic spectra with the observations, we decided to consider only the transmission spectra obtained by \citet{cauley2019} and \citet{wyttenbach2020}. This is because \citet{yan2018} and \citet{turner2020} provided only the \ha\ line, the profile of \citet{turner2020} is much noisier than the others, and the profile of \citet{yan2018} appears to be significantly different (deeper and broader) from the others. To account for the differences in the \ha\ and \hb\ line profiles and to increase the signal-to-noise of the transmission spectra, we aligned the profiles of \citet{cauley2019} and \citet{wyttenbach2020} employing the line centers obtained from the Gaussian fits and computed the mean profiles using a weighted average. Therefore, in the end, the comparison with synthetic spectra is based on the profiles obtained by \citet{cauley2019} and \citet{wyttenbach2020}, and on the weighted average profiles obtained considering the results of \citet{cauley2019} and \citet{wyttenbach2020}. 
\begin{table*}[ht!]
\caption{Line depth, FWHM, and equivalent width derived from the Gaussian fits to the line profiles of \citet{yan2018}, \citet{cauley2019}, \citet{turner2020}, \citet{wyttenbach2020}, and of the weighted average line profile obtained averaging those of \citet{cauley2019} and \citet{wyttenbach2020}.}
\label{tab:observations}
\begin{center}
\begin{tabular}{l|ccc|ccc}
\hline
\hline
Author & \multicolumn{3}{c|}{H$\alpha$} & \multicolumn{3}{c}{H$\beta$} \\
Author & Depth [\%] & FWHM [\AA] & Equivalent   & Depth [\%] & FWHM [\AA] & Equivalent   \\
       &            &            & width [m\AA] &            &            & width [m\AA] \\
\hline
\citet{yan2018}        & 1.13$\pm$0.02 &  1.13$\pm$0.01  & 13.5$\pm$0.2  &  $-$ & $-$ & $-$ \\
\citet{cauley2019}     & 1.13$\pm$0.03 &  0.80$\pm$0.01  &  9.4$\pm$0.2  &  0.83$\pm$0.04 & 0.55$\pm$0.01 & 5.0$\pm$0.2 \\
\citet{turner2020}     & 1.05$\pm$0.10 &  0.85$\pm$0.04  &  8.8$\pm$0.6  &  $-$ & $-$ & $-$ \\
\citet{wyttenbach2020} & 0.92$\pm$0.03 &  0.97$\pm$0.02  &  9.4$\pm$0.2  &  0.68$\pm$0.04 & 0.74$\pm$0.02 & 5.1$\pm$0.2 \\
Weighted average       & 1.07$\pm$0.02 & 0.850$\pm$0.008 &  9.4$\pm$0.1  &  0.77$\pm$0.03 & 0.64$\pm$0.01 & 5.0$\pm$0.1 \\
\hline
\end{tabular}
\end{center}
\end{table*}
%
\section{Temperature-pressure profiles}\label{sec:tp}
\begin{figure*}[ht!]
\includegraphics[width=\hsize,clip]{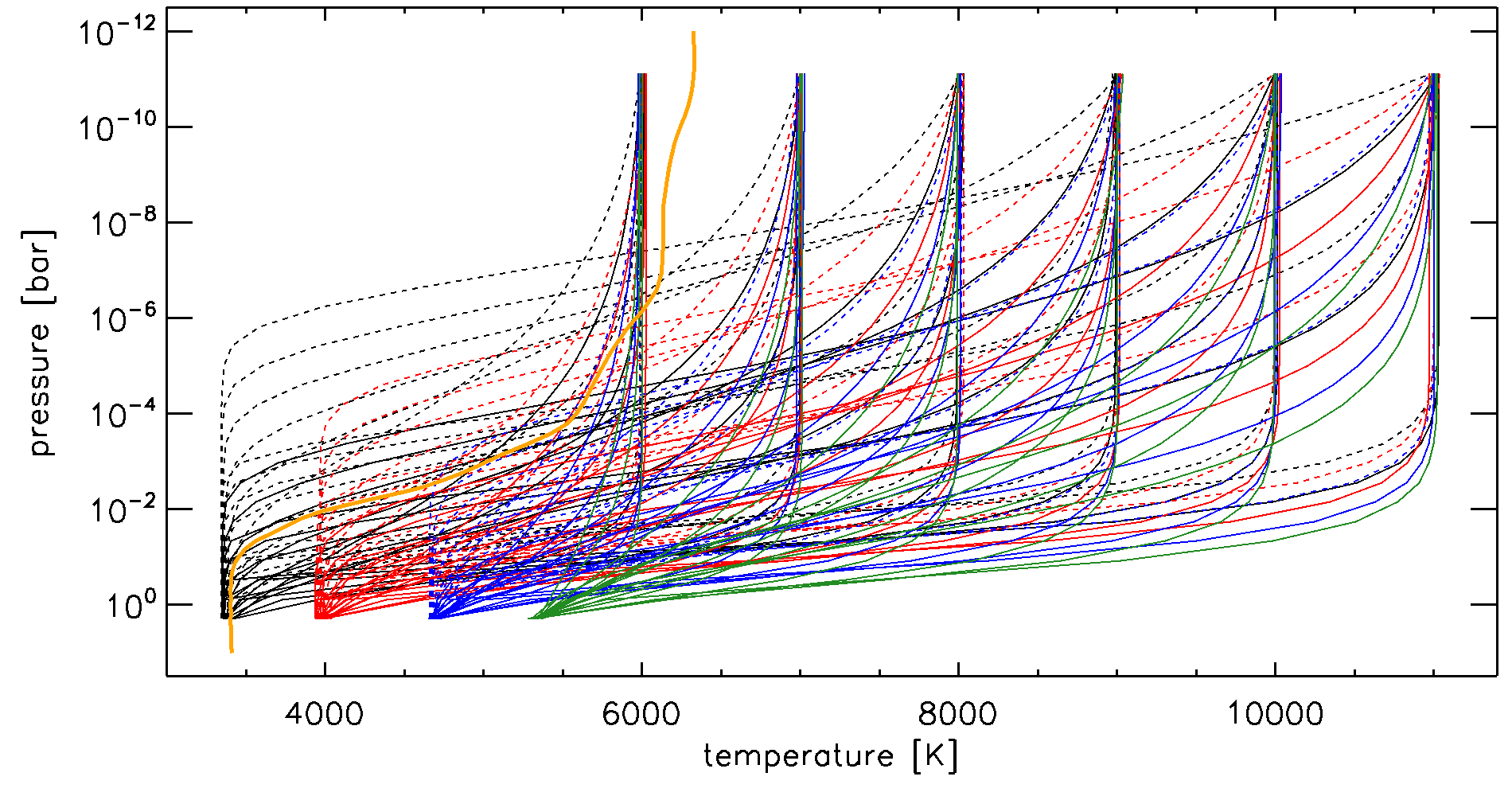}
\caption{Empirical TP profiles used as input for the spectral synthesis calculations. Black, red, blue, and green lines identify the models having a lower temperature, at the bottom of the atmosphere, close to 3500, 4000, 4750, and 5500\,K, respectively. Solid lines are for profiles computed setting $\kappa$ equal to 2.5, while dashed lines are for profiles computed setting $\kappa$ equal to 3.1 The thick orange line shows the TP profile obtained for KELT-9b using PHOENIX.}
\label{fig:tp_input}
\end{figure*}

\citet{lothringer2018} employed the PHOENIX stellar and planetary atmosphere code to compute TP profiles of ultra-hot Jupiters showing that they present strong temperature inversions in the middle part of the atmosphere, at pressures between 10$^{-1}$ and 10$^{-5}$\,bar, caused by absorption of the intense UV and optical stellar radiation \citep[see also][]{garcia2019}. \citet{fossati2018b} used a PHOENIX TP profile, specifically computed for KELT-9b, to estimate the planetary mass-loss rate. Also this profile shows a rather steep temperature inversion, with the increase occurring between 10$^{-1}$ and 10$^{-6}$\,bar (Figure~\ref{fig:tp_input}). The presence of an inverted temperature profile has been observationally confirmed by \citet{pino2020}. 

Therefore, we took inspiration from the presence of the inversion around the mbar level to draw a large number of empirical TP profiles, which are then used as basis to compute the synthetic transmission spectra (Sect.~\ref{sec:transmission}) to be compared to the observations (Sect.~\ref{sec:results}). We computed 126 empirical TP profiles using a modified version of the Three-channel Eddington Approximation (TCEA) temperature profile model by \citet{guillot2010}, in the form used by \citet{line2013}. In this approximation, the temperature profile ($T$) as a function of the atmospheric pressure ($p$) is parametrised as
\begin{equation}
\label{eq.Tp}
T(p) = \left\{0.75\,\left[\left(\frac{2}{3}+\tau\right)\,t_{\rm int}^4+\xi\,t_{\rm irr}^4\right]\right\}^{0.25}\,,
\end{equation}
where $t_{\rm int}$ is the planetary internal temperature,
\begin{equation}
\label{eq.tau}
\tau = \frac{\kappa\,10^{\frac{p}{s}}}{g_{\rm p}}\,,
\end{equation}
\begin{equation}
\label{eq.xi}
\xi = \frac{2}{3}\,\left\{\frac{1}{\gamma}\,\left[1+\left(0.5\gamma\tau-1\right)\,e^{-\gamma\tau}\right]+\gamma\left(1-0.5\tau^2\right)E_2(\gamma\tau)+1\right\}\,,
\end{equation}
and
\begin{equation}
\label{eq.tirr}
t_{\rm irr} = \beta\,\sqrt{\frac{R_{\rm s}}{2a}}\,T_{\rm eff}\,.
\vspace{0.02 cm}
\end{equation}
The parameters of the model that we vary to modify the shape of the TP profile are $\kappa$, $\gamma$, $\beta$, and $s$. The parameter $\kappa$, which is related to the Planck thermal infrared opacity, shifts the TP profile in pressure. The parameter $\gamma$, which is related to the visible-to-thermal stream Planck mean opacity ratio, controls the difference in temperature between the top and bottom of the atmosphere. The parameter $\beta$, which is related to various atmospheric temperature distribution effects (e.g., albedo, emissivity, day-night redistribution), shifts the TP profile in temperature. The parameter $s$ was not present in the original formalism of \citet{guillot2010} and \citet{line2013} and we introduce it to control the slope of the temperature gradient in the atmosphere. To reduce the number of free parameters, we set $t_{\rm int}$ equal to zero, hence Eq.~(\ref{eq.Tp}) becomes
\begin{equation}
\label{eq.Tp_short}
T(p) = \left(0.75\,\xi\,t_{\rm irr}^4\right)^{0.25}\,=\,\left(0.75\,\xi\right)^{0.25}\,t_{\rm irr}\,.
\end{equation}
In Eq.~(\ref{eq.tau}), $g_{\rm p}$ is the planetary surface gravity computed considering the measured planetary mass and transit radius and, in Eq.~(\ref{eq.xi}), $E_2(\gamma\tau)$ is the second order exponential integral function of $\gamma\tau$. In Eq.~(\ref{eq.tirr}), $R_{\rm s}$ is the stellar radius, $a$ is the planetary orbital separation, and $T_{\rm eff}$ is the stellar effective temperature. Table~\ref{tab.system_parameters} lists the relevant system parameters adopted in this work.
\begin{table}[ht!]
\caption{Adopted KELT-9 system parameters.}
\label{tab.system_parameters}
\begin{center}
\begin{tabular}{l|cc}
\hline
\hline
Parameter & Value & Source \\
\hline
$T_{\rm eff}$ [K]           & 9600    & \citet{borsa2019} \\
$M_{\rm s}$ [$M_{\odot}$]   & 2.32    & \citet{borsa2019} \\
$R_{\rm s}$ [$R_{\odot}$]   & 2.418   & \citet{borsa2019} \\
$a$ [AU]                    & 0.03368 & \citet{borsa2019} \\
$M_{\rm p}$ [$M_{\rm Jup}$] & 2.88    & \citet{borsa2019} \\
$R_{\rm p}$ [$R_{\rm Jup}$] & 1.936   & \citet{borsa2019} \\
$b$                         & 0.168   & \citet{borsa2019} \\
\hline
\end{tabular}
\end{center}
\end{table}

The TP profiles have been constructed considering the already available constraints. For setting the range of temperatures spanned by the models in the lower atmosphere, we considered the day- and night-side temperatures measured from phase curve observations \citep{mansfield2020,wong2019}. We further considered that the TP profile should present a temperature rise \citep{lothringer2018,pino2020} and that the temperature in the PHOENIX TP profile at pressures below about 10\,mbar may be underestimated \citep{turner2020}. At the bottom of the atmosphere, around the 1\,bar level, the empirical TP profiles have temperatures ranging between about 3500 and 5500\,K (i.e., about 1000\,K above and below the measured day-side temperature), while at the top of the atmosphere the TP profiles have temperatures ranging between about 6000 and 11000\,K. We did not consider higher upper atmospheric temperatures because it would be unlikely for the planet to have an upper atmosphere hotter than the stellar photosphere (see Table~\ref{tab.system_parameters}), which is the source of heating \citep[see also][]{mitani2020}. Giant planets in close orbit to late-type stars present an upper atmospheric temperature hotter than the stellar photosphere, because the heating source is the radiation emitted by the stellar chromosphere and transition region, which are significantly hotter than the stellar photosphere. However, this is not the case for KELT-9b because the star does not possess a chromosphere and transition region \citep{fossati2018b}. Therefore, in the absence of an atmospheric heating mechanism other than photospheric stellar irradiation, an atmospheric temperature significantly higher than that of the stellar photosphere would be unlikely.

The 126 empirical TP profiles are divided into families having four different minimum temperatures at the bottom of the atmosphere (i.e., about 3500, 4000, 4750, and 5500\,K) and six different maximum temperatures at the top of the atmosphere (i.e., about 6000, 7000, 8000, 9000, 10000, and 11000\,K). In addition to the temperature minima and maxima, we varied also the pressure level at which the temperature rises and the slope of the temperature increase. We computed the TP profiles by setting $\kappa$ equal to 2.5 or 3.1, $\gamma$ ranging between 0.28 and 2.32 (81 different values), $\beta$ ranging between 0.98 and 1.62 (34 different values), and $s$ equal to 1, 2.5, or 4. For each TP profile, we set the maximum pressure (i.e., bottom of the atmosphere) at 2\,bar and the minimum pressure (i.e., top of the atmosphere) at 8$\times$10$^{-12}$\,bar, and divided the atmosphere into 29 layers equally spaced in $\log{p}$ (i.e., steps of 0.408 in $\log{p}$). This number of layers is a compromise between describing the TP profiles with enough accuracy and computation time. In this respect, for a few TP profiles, we run additional models with a larger number of layers (up to 35) obtaining the same results. This range of pressures is wide enough to fully contain the atmospheric formation region of the hydrogen Balmer lines \citep{turner2020} and ensures that the atmosphere is completely transparent to optical light at the top of the atmosphere and completely opaque to optical light at the bottom of the atmosphere. Figure~\ref{fig:tp_input} shows all TP profiles, also comparing them with that obtained with PHOENIX for KELT-9b, while Table~\ref{tab.TPprofiles} lists the TP profiles and the parameters employed to obtain them. For reference, among our set of TP profiles, that computed by PHOENIX for KELT-9b is best reproduced by the TP profile number 025, which has $\kappa$ equal to 3.1, $\gamma$ equal to 0.99, $\beta$ equal to 1.03, and $s$ equal to 2.5. As mentioned above, the majority of the TP profiles is on average hotter than that computed by PHOENIX, because this is what was suggested by the analyses of \citet{garcia2019}, \citet{turner2020}, and \citet{wyttenbach2020}.
%
\section{Synthetic transmission spectra}\label{sec:transmission}
To compute the synthetic \ha\ and \hb\ transmission spectra on the basis of the TP profiles shown in Fig.~\ref{fig:tp_input}, we employed the Cloudy spectral synthesis code following the scheme described in \citet{turner2020} and \citet{young2020}. Cloudy is a spectral synthesis code designed to simulate physical conditions within an astrophysical plasma, predicting the emitted/absorbed spectrum, further accounting for NLTE effects \citep{ferland2017}, which is important for correctly computing the population of atoms in excited states, such as those leading to the formation of the hydrogen Balmer lines \citep[e.g.,][]{garcia2019,young2020}. Cloudy further accounts for hydrogen and metal collisional excitation and ionisation, as well as photexcitation and photoionisation.

Since our empirical TP profiles are one-dimensional, we assume a spherically symmetric planetary atmosphere varying only with altitude. Cloudy is capable to consider all elements up to Zn and, to speed up the calculations, we took into account just H, He, C, O, Na, Mg, Si, K, Ca, Ti, and Fe, which are abundant and provide a significant contribution to the electron density through their ionisation, hence also to the mean molecular weight and continuum level. The only molecules we took into account are those of hydrogen, namely H$_2$, H$_2^+$, and H$_3^+$ \citep[see][for more details]{young2020}. Other molecules play a minor role in determining the atmospheric physical properties \citep{lothringer2019}, particularly around the formation region of the hydrogen Balmer lines. We took the atmospheric abundance profile of the considered elements from the PHOENIX model, which assumes solar metallicity, and let Cloudy compute dissociation, ionisation, and excitation. 

In the atmospheric temperature profiles, the radial dimension is given by the pressure, but the Cloudy calculations require a physical length. Therefore, we computed the radius of each layer $i$ using the pressure scale height
\begin{equation}
\label{eq.scale_height}
H_i = \frac{k_B\,T_i}{\mu_i\,g_i}\,,
\end{equation}
where $k_B$ is the Boltzmann constant, $T_i$ is the temperature of layer $i$, $\mu_i$ is the mean molecular weight of layer $i$, and $g_i$ is the planetary gravity at layer $i$. We then obtained the planetary radius corresponding to each layer by computing
\begin{equation}
\label{eq.ri}
r_i = 
\begin{cases}
-H_i\,\ln{(p_i/p_{i-1})} + r_{i-1} & p_i > p_0 \\
-H_i\,\ln{(p_i/p_{i+1})} + r_{i+1} & p_i \leq p_0
\end{cases}
\end{equation}
where $p_i$ is the atmospheric pressure at layer $i$ and $p_0$ is the reference pressure at the observed planetary radius $R_{\rm p}$. 

Because of the short planetary orbital distance and rather high stellar mass, the Roche lobe of KELT-9b is smaller than that of most hot Jupiters and it plays a significant role in shaping the atmospheric properties \citep[e.g.,][]{fossati2018b}. For this reason, when computing $g_i$ in Eq.~(\ref{eq.scale_height}), we accounted for the shape and location of the Roche lobe in the direction perpendicular to the sub-stellar point, i.e. that probed by transmission spectroscopy.

We mapped each one-dimensional (1D) TP profile onto concentric circles and then calculated the lengths through successive layers of atmosphere along line-of-sight transmission chords \citep[see][]{turner2020,young2020}. These lengths, along with the atmospheric properties of their respective layers, have been stacked and entered into Cloudy as the line of sight transmission medium. We then computed separate transmission spectra with Cloudy for each layer, at a spectral resolution of $R$\,=\,100,000 and without adding any turbulent velocity. Finally, we computed the total transmission spectrum of the planet by adding up the single layer spectra, weighted by their relative area projected on the stellar disc and accounting for the planetary impact parameter ($b$). When doing this last operation, we did not account for limb darkening effects because the observations of \citet{yan2018}, \citet{cauley2019}, and \citet{wyttenbach2020}, which are those with the higher signal-to-noise ratio and hence more relevant for the analysis, are already corrected for the CLV effect. Further details on the algorithm employed to generate transmission spectra with Cloudy are given by \citet{young2020}. Because of the duration of Cloudy calculations, it is not possible to carry out this work accounting for the three dimensional geometry of the atmosphere, but we discuss the impact of the 1D assumption in Sect.~\ref{sec:assumptions}. Overall, the broadening terms considered in computing the synthetic profiles are natural, temperature, Stark, and instrumental broadening. We run a few models testing the relative impact of the different broadening terms (excluding instrumental broadening) obtaining that thermal broadening is the most relevant, while Stark broadening is the least important. We do not consider rotational broadening, which would however have a small impact on the line profiles, compared in particular to thermal broadening.

Following what indicated by the PHOENIX model, we first took $p_0$ to be equal to 0.01\,bar \citep{turner2020}. However, because the transmission spectra are at the end normalised to the continuum, the choice of the reference pressure affects the line strength of the synthetic spectra. Furthermore, because each synthetic spectrum is computed employing a different TP profile, the reference pressure depends on the TP profile. To identify the reference pressure to employ to compute each transmission spectrum, we applied an iterative procedure. We first computed one transmission spectrum for each TP profile assuming $p_0$\,=\,0.01\,bar. Then, for each transmission spectrum, we set $p_0$ equal to the pressure value of the layer for which the average continuum opacity calculated by Cloudy in the region covered by the \ha\ and \hb\ lines is closest to 2/3 (the continuum opacity at the wavelengths covered by the \ha\ and \hb\ lines is due to H$^-$). Finally, we recomputed a new transmission spectrum with the updated reference pressure and followed this procedure again, until convergence. We reached convergence for all profiles following at most four iterations. The Cloudy code is not parallel, therefore we parallelised the computation of the different transmission rays. Employing 126 CPUs, split into two clusters to avoid memory overflow, we computed the transmission spectra, including the convergence to obtain the continuum level, within about 2.5 weeks.

\begin{figure}[h!]
\includegraphics[width=\hsize,clip]{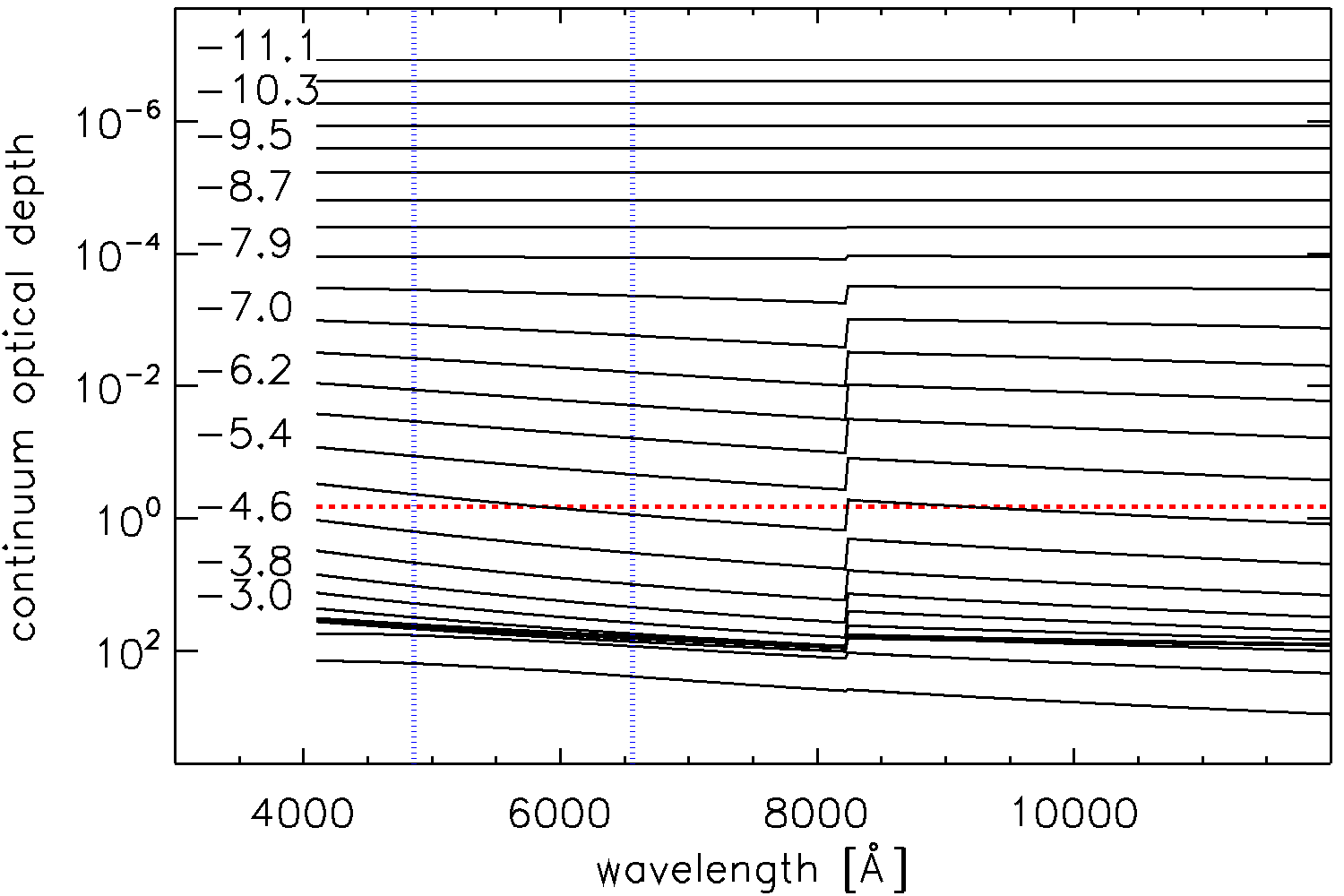}
\caption{Continuum optical depth as a function of wavelength in the optical range for each atmospheric layer in transmission geometry considering the TP model number 047. The values shown on the left-hand side give the logarithm of the pressure of the corresponding layer. For visualisation purposes, the pressure level is shown for every second layer starting from the top of the atmosphere and stop where the continuum optical depths of the layers start overlapping. To guide the eye, the red horizontal line is at 2/3, while the vertical blue dotted lines indicate the position of the \ha\ and \hb\ lines.}
\label{fig:continuum}
\end{figure}

Figure~\ref{fig:continuum} shows, as an example, the continuum optical depth as a function of wavelength for each atmospheric layer computed with Cloudy in transmission geometry at the last iteration considering the TP model number 047. The continuum optical depth around the position of the \ha\ and \hb\ lines is closest to 2/3 at the level corresponding to a pressure of 10\,$\mu$bar.
%
\section{Results}\label{sec:results}
Before comparing the observed and synthetic line profiles, we normalised the synthetic spectra by fitting first order polynomials (one for the \ha\ and one for the \hb\ line) to continuum points adjacent to and on both sides of each line. We also corrected the observations for wavelength shifts employing the line centers obtained from the Gaussian fits. 

We compared the observed and synthetic profiles employing the $\chi^2$ and the line equivalent widths as diagnostics. For the former, $\chi^2$ has been computed employing the non-rebinned observed spectra and by interpolating the synthetic spectra on the sampling of each observation. For the latter, we compared the observed and synthetic equivalent widths and also the ratio of the \ha-to-\hb\ equivalent widths. Both $\chi^2$ and equivalent widths have been evaluated considering a range around the line center of 2\,\AA\ for \ha\ and of 1.4\,\AA\ for \hb. In the case of the \ha\ line there are 141 degrees of freedom, while in the case of the \hb\ line there are 133 degrees of freedom. 
\begin{figure*}[ht!]
\includegraphics[width=\hsize,clip]{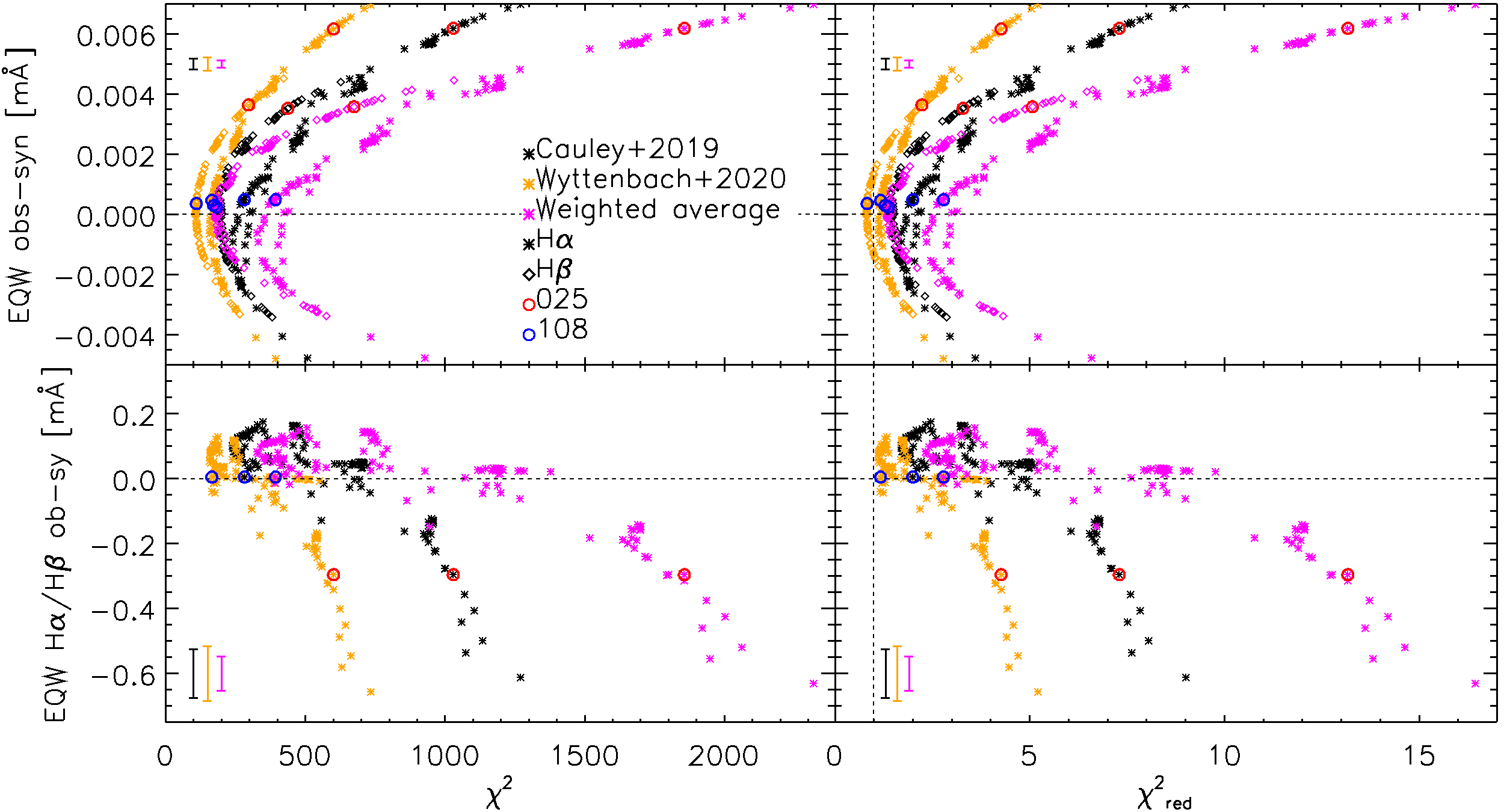}
\caption{Top-left: $\chi^2$ vs difference between observed and modelled equivalent width (EQW) considering the profiles obtained by \citet[][black]{cauley2019}, \citet[][orange]{wyttenbach2020}, and for the weighted average (purple). Asterisks and rhombi are for the \ha\ and \hb\ profiles, respectively. Bottom-left: $\chi^2$ obtained from the analysis of the \ha\ line profiles vs difference between observed and modelled \ha-to-\hb\ equivalent width ratio. The error bars at the top-left (top panel) and bottom-left (bottom panel) corners indicate the size of the equivalent width and equivalent width ratio uncertainties. Right: same as left, but with $\chi^2_{\rm red}$ on the x-axis. The dashed lines at zero and one are to guide the eye. In all panels, the position of the results obtained with the TP profiles number 025 and 108 is marked by red and blue circles, respectively.}
\label{fig:EQWandCHI2solar}
\end{figure*}

We consider both $\chi^2$ and equivalent widths, instead of just $\chi^2$, to reduce the importance of reproducing line shape over line strength. Reproducing line profiles implies fitting both strength and shape of the lines, including line widths and shape of the wings, which is significantly more challenging than reproducing line strengths alone, that is equivalent widths. The widths and wings shape of spectral lines are influenced by several factors (e.g., turbulence) and physical processes (e.g., collisions), which are less likely to be correctly reproduced by the modelling. Instead, line strengths are more influenced by the general physical properties of the gas, such as temperature and densities, which are those we primarily aim to characterise.

Figure~\ref{fig:EQWandCHI2solar} shows the results obtained from the $\chi^2$ analysis and from the comparison of the equivalent widths. The top-left panel of Fig.~\ref{fig:EQWandCHI2solar} indicates that there is a family of TP profiles that both minimises $\chi^2$, reaching values of $\chi^2$ between 100 and 400 (i.e., reduced $\chi^2$ -- $\chi^2_{\rm red}$ -- between about 1 and 3; see top-right panel of Fig.~\ref{fig:EQWandCHI2solar}) for both lines, and matches well the measured equivalent widths. However, the bottom panels of Fig.~\ref{fig:EQWandCHI2solar} show that this family of models does not fit equally well the measured \ha-to-\hb\ equivalent width ratio.

In general, because of the smaller error bars, the $\chi^2$ obtained from the analysis of the weighted average profiles is larger than that obtained from the analysis of the single profiles. We also find that the profiles of \citet{wyttenbach2020} are a better match to the synthetic lines. This is because the \ha\ and \hb\ lines obtained by \citet{wyttenbach2020} are shallower and broader than those of \citet{cauley2019}. We come back to this point later in this Section.

\begin{figure}[h!]
\includegraphics[width=\hsize,clip]{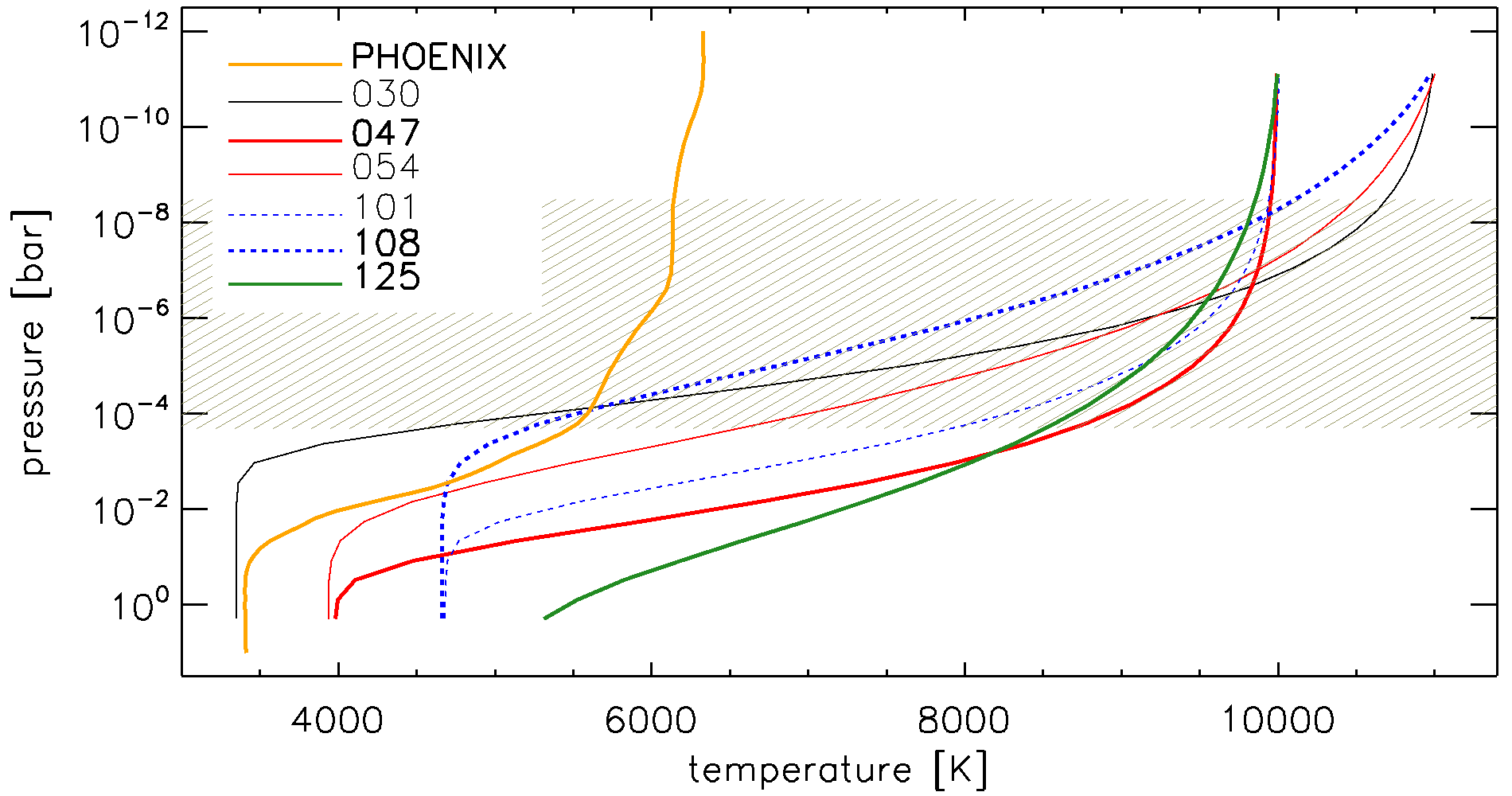}
\caption{Same as Fig.~\ref{fig:tp_input}, but for the TP profiles best fitting the observed \ha\ and \hb\ lines according to the conditions listed in Eq.~(\ref{eq:conditions}). The thicker lines indicate the three TP profiles fulfilling the stricter conditions listed in Eq.~(\ref{eq:conditions_strict}). The hatched area shows the main formation region of the \ha\ and \hb\ lines according to the three best fitting models.}
\label{fig:TPbestfits}
\end{figure}

We identified the family of TP profiles best fitting the observations on the basis of the considered diagnostics by extracting those fulfilling the conditions
\begin{equation}
\label{eq:conditions}
\begin{array}{l}
\,\chi^2_{\rm red,H\alpha,WA} \leq 3.0 \\
\,\chi^2_{\rm red,H\beta,WA} \leq 1.8 \\
\,|\Delta\,EQW_{\rm H\alpha,WA}| \leq 0.0012 \\
\,|\Delta\,EQW_{\rm H\beta,WA}| \leq 0.0012 \\
\,|\Delta\,EQW_{\rm ratio,WA}| \leq 0.12\,,
\end{array}
\end{equation}
where $\chi^2_{\rm red,H\alpha,WA}$ and $\chi^2_{\rm red,H\beta,WA}$ are the $\chi^2_{\rm red}$ values computed for the weighted average \ha\ and \hb\ line profiles, respectively, $|\Delta\,EQW_{\rm H\alpha,WA}|$ and $|\Delta\,EQW_{\rm H\beta,WA}|$ are the absolute values of the difference between observed and synthetic equivalent widths considering the weighted average \ha\ and \hb\ line profiles, respectively, and $|\Delta\,EQW_{\rm ratio,WA}|$ is the absolute value of the difference between observed and modelled \ha-to-\hb\ equivalent width ratio computed considering the weighted average line profiles. Under these conditions, the best fitting models correspond to those of the TP profiles number 030, 047, 054, 101, 108, and 125 that are shown in Fig.~\ref{fig:TPbestfits}. We further looked for the three best fitting models by setting more stringent conditions, namely
\begin{equation}
\label{eq:conditions_strict}
\begin{array}{l}
\,\chi^2_{\rm red,H\alpha,WA} \leq 2.8 \\
\,\chi^2_{\rm red,H\beta,WA} \leq 1.6 \\
\,|\Delta\,EQW_{\rm H\alpha,WA}| \leq 0.0011 \\
\,|\Delta\,EQW_{\rm H\beta,WA}| \leq 0.0011 \\
\,|\Delta\,EQW_{\rm ratio,WA}| \leq 0.11\,,
\end{array}
\end{equation}
obtaining that the best fitting TP profiles are 047, 108, and 125, which are highlighted in Fig.~\ref{fig:TPbestfits} by thicker lines. The values given in the conditions listed in Eq.~(\ref{eq:conditions}) and (\ref{eq:conditions_strict}) are arbitrary. This is because we just aim at identifying the general properties of the family of TP profiles leading to best fit the observations. In the next section, we then focus as an example on just one of those TP profiles, which has been chosen on the basis of considerations independent from the fitting of the hydrogen Balmer lines.

Figure~\ref{fig:TPbestfits} clearly shows that the observations are best reproduced by models based on TP profiles having an upper atmospheric temperature around 10000--11000\,K, namely about 4000\,K hotter than what predicted by the PHOENIX model and comparable to the stellar effective temperature. In general, synthetic lines computed on the basis of TP profiles with larger $\gamma$ values are a better fit to the observations. Furthermore, the family of TP profiles best fitting the observations presents the temperature inversion in the middle of the atmosphere, roughly between 10\,mbar and 1\,$\mu$bar, indicating that the best fitting TP profiles are neither those with a very steep inversion at the bottom of the atmosphere, nor those with the inversion high up in the atmosphere. Figure~\ref{fig:TPbestfits} also suggests that the observed \ha\ and \hb\ line profiles do not depend of the temperature of the lower atmosphere (i.e., $\gtrsim$0.1\,bar). However, the planetary temperature measured through phase curve observations can be at aid in further identifying the best fitting TP profile. As a matter of fact, at the photosphere ($\approx$10\,mbar level), the TP profiles number 047 and 125 are significantly hotter than the measured planetary day-side temperature, making TP model number 108 the most likely of the three \citep[see also][]{lothringer2020}. Therefore, the TP profile number 108 is the one we employ to represent the family of TP profiles best fitting the observations and that we thoroughly discuss in the next section.

We performed this same analysis again, but considering the profiles of \citet{cauley2019} and \citet{wyttenbach2020} separately. We obtained that the result on the temperature of the upper atmosphere is robust as is does not depend on the chosen dataset, while the result on the shape and position of the temperature inversion depends slightly on the considered dataset, with the profiles obtained by \citet{cauley2019} leading to a steeper temperature inversion located deeper in the atmosphere compared to what we obtained from the analysis of the profiles given by \citet{wyttenbach2020}.

\begin{figure*}[ht!]
\includegraphics[width=\hsize,clip]{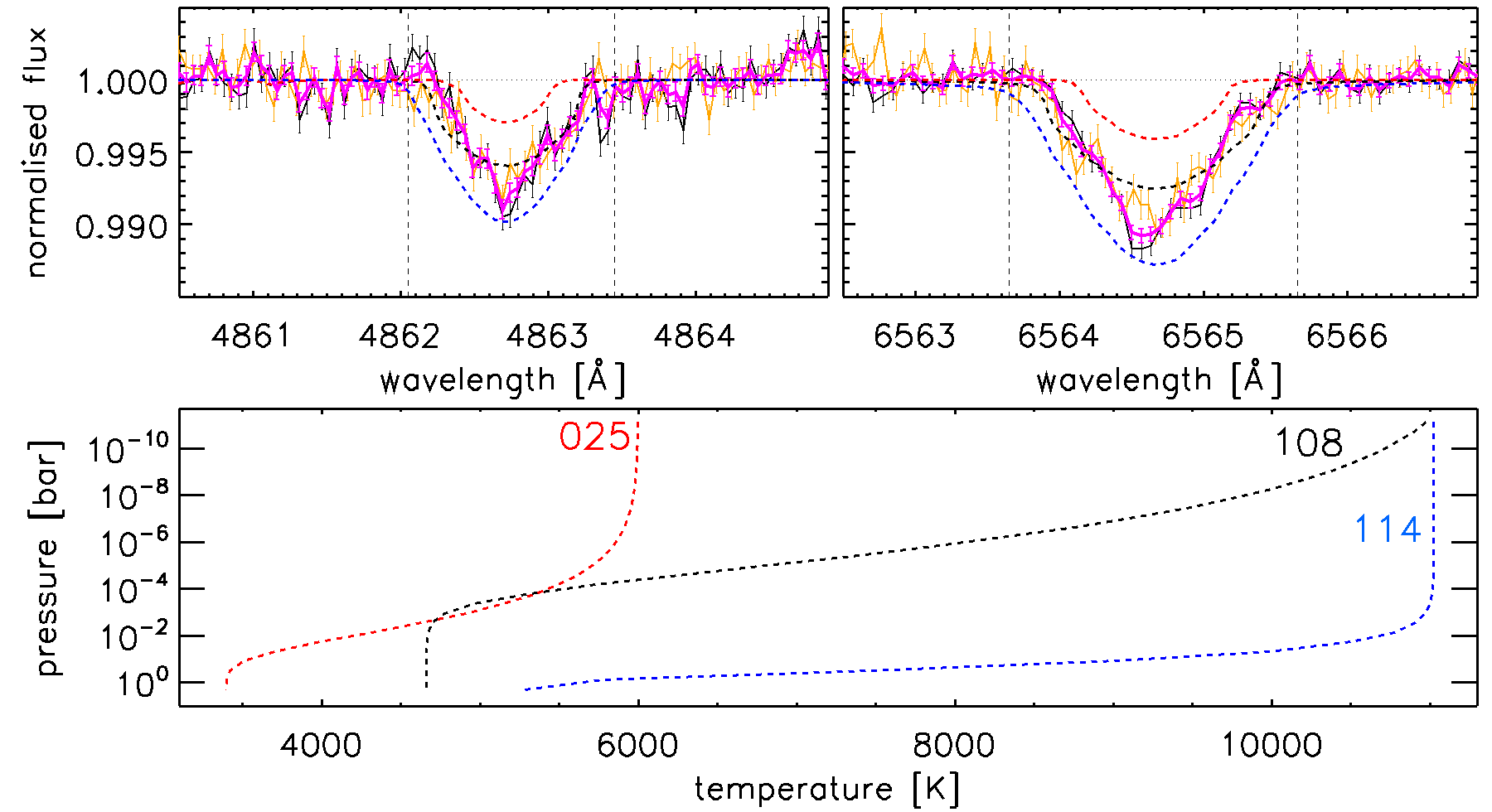}
\caption{Top: same as Fig.~\ref{fig:comparison_observations}, but the dashed black, red, and blue lines are the hydrogen Balmer line profiles obtained employing the TP profiles number 108, 025, and 114, respectively. The TP profile number 025 is the most similar to the PHOENIX self-consistent TP profile, the TP profile number 108 is the one leading to best fit the observed hydrogen Balmer lines, while the TP profile number 114 is the one leading to generate the strongest Balmer lines within our sample. The black, vertical dashed lines enclose the wavelength ranges considered to compute $\chi^2$ and the equivalent widths. Bottom: TP profiles corresponding to the 108 (black), 025 (red), and 114 (blue) model numbers.}
\label{fig:profilesANDtp}
\end{figure*}

Figure~\ref{fig:profilesANDtp} shows the observed \ha\ and \hb\ line profiles in comparison with the synthetic lines obtained employing the TP profiles number 108 (thick red solid line in Fig.~\ref{fig:TPbestfits}), 025, which is the TP profile most resembling the PHOENIX one, and 114, which is the TP profile leading to the strongest Balmer lines. The best fitting profiles are slightly broader and shallower than the weighted averaged \ha\ and \hb\ lines. The line depths of the best fitting synthetic profiles match well those of the profiles given by \citet{wyttenbach2020}, but are broader, while they are shallower and even broader than those of \citet{cauley2019}.

In agreement with \citet{turner2020}, the synthetic profile obtained employing a TP structure comparable to that given by PHOENIX leads to significantly weaker lines. The synthetic spectra presenting the strongest \ha\ and \hb\ absorption features are only slightly deeper than those of \citet{cauley2019}, but are significantly broader, which is why both equivalent width and $\chi^2_{\rm red}$ analyses do not favour these models.

\begin{figure}[h!]
\includegraphics[width=\hsize,clip]{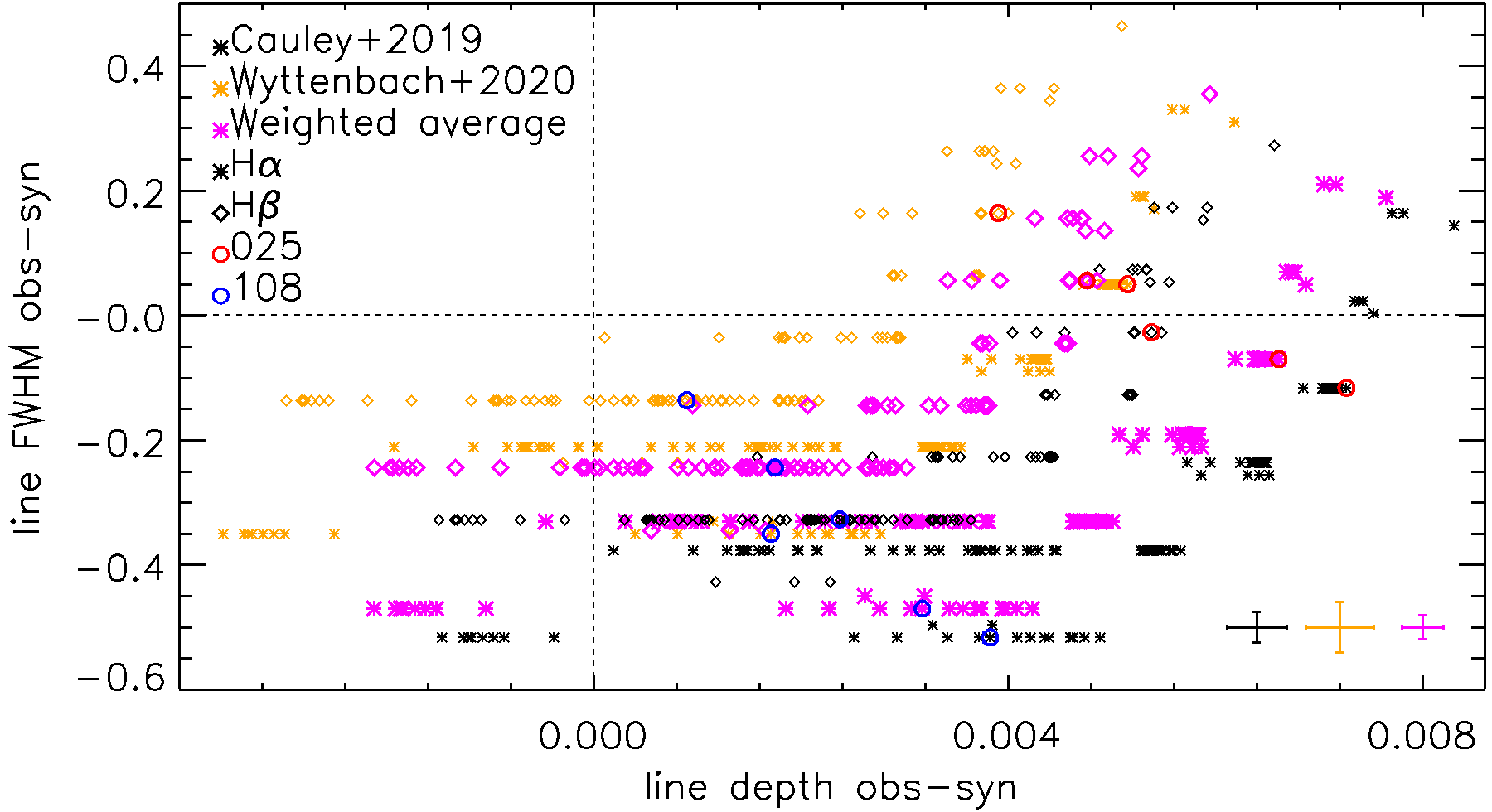}
\caption{Difference between observed and synthetic line FWHM as a function of the difference between observed and synthetic line depth for the profiles obtained by \citet[][black]{cauley2019}, \citet[][orange]{wyttenbach2020}, and for the weighted average profile (purple). Asterisks and rhombi are for the \ha\ and \hb\ lines, respectively. The dashed horizontal and vertical lines at 0 are to guide the eye. The models best fitting the observed line depths overestimate the lines FWHM, and vice-versa. The error bars at the bottom-right corner indicate the size of the uncertainties on the measured line depth and FWHM. The
position of the results obtained with the 025 and 108 TP profiles is marked by red and blue circles, respectively.}
\label{fig:DEPTHvsWIDTH}
\end{figure}

In general, as shown in Fig.~\ref{fig:DEPTHvsWIDTH}, the synthetic profiles that best fit the observed line depths, overestimate the observed widths. We remark that the broadening terms considered in computing the synthetic profiles are natural, temperature, Stark, and instrumental broadening, and that Stark broadening has a negligible impact on the total line broadening. It may be possible that the data analysis procedure, particularly the spectral normalisation, removes the signal coming from the far line wings, which the synthetic spectra suggest being not negligible, particularly for \ha, hence artificially reducing the width of the observed lines. 
%
\section{Discussion}\label{sec:discussion}
We analyse here in detail the results. We focus on the TP model profiles number 025, which is the closest to the one obtained with PHOENIX, and number 108, which, of the three models best fitting the planetary H$\alpha$ and H$\beta$ line profiles, is the one with the temperature of the lower atmosphere closer to what obtained from phase curve observations \citep{wong2019,mansfield2020,lothringer2020}.
\subsection{Atmospheric composition}\label{sec:Hcomposition}
\begin{figure}[h!]
\includegraphics[width=\hsize,clip]{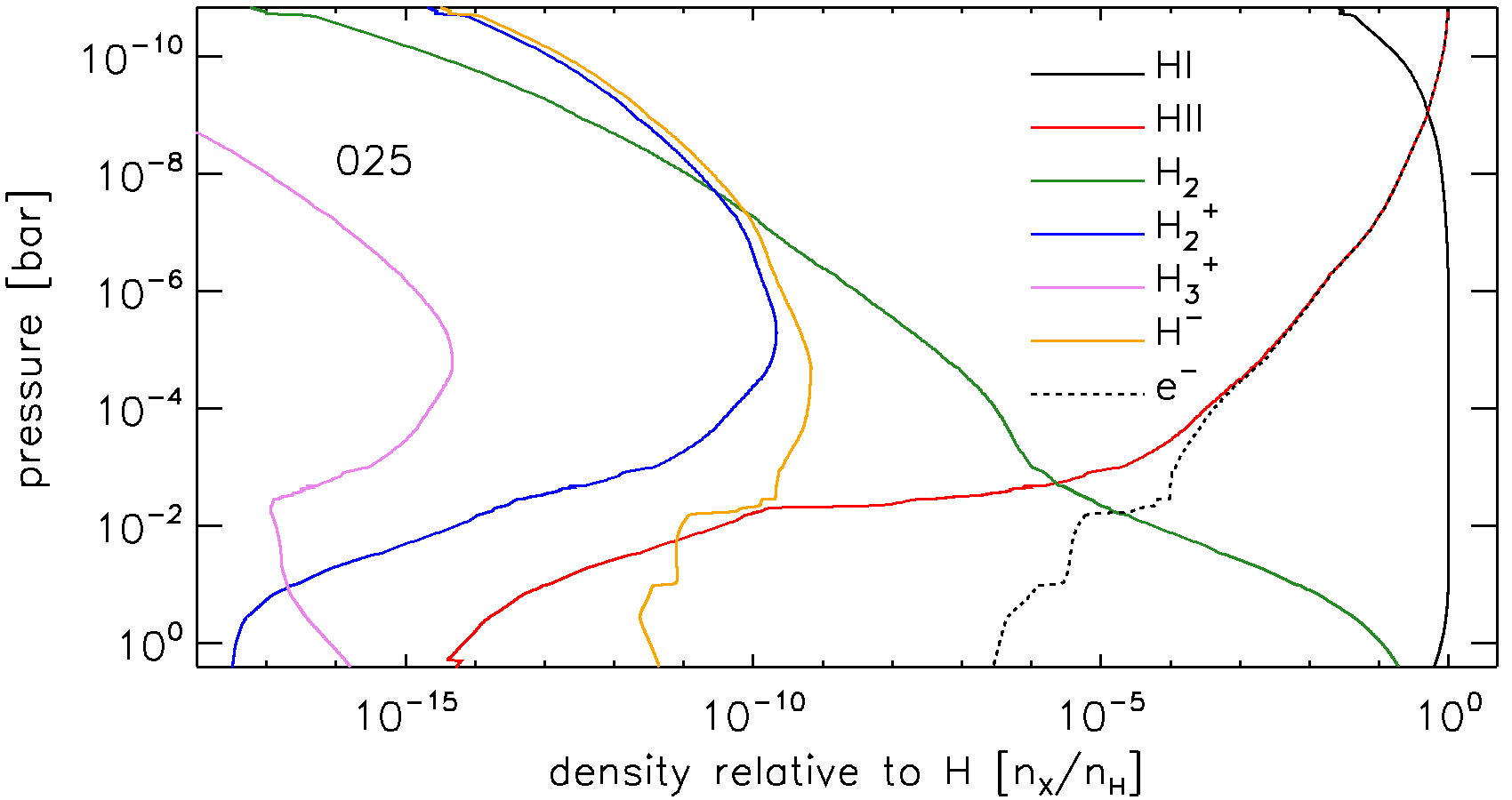}
\includegraphics[width=\hsize,clip]{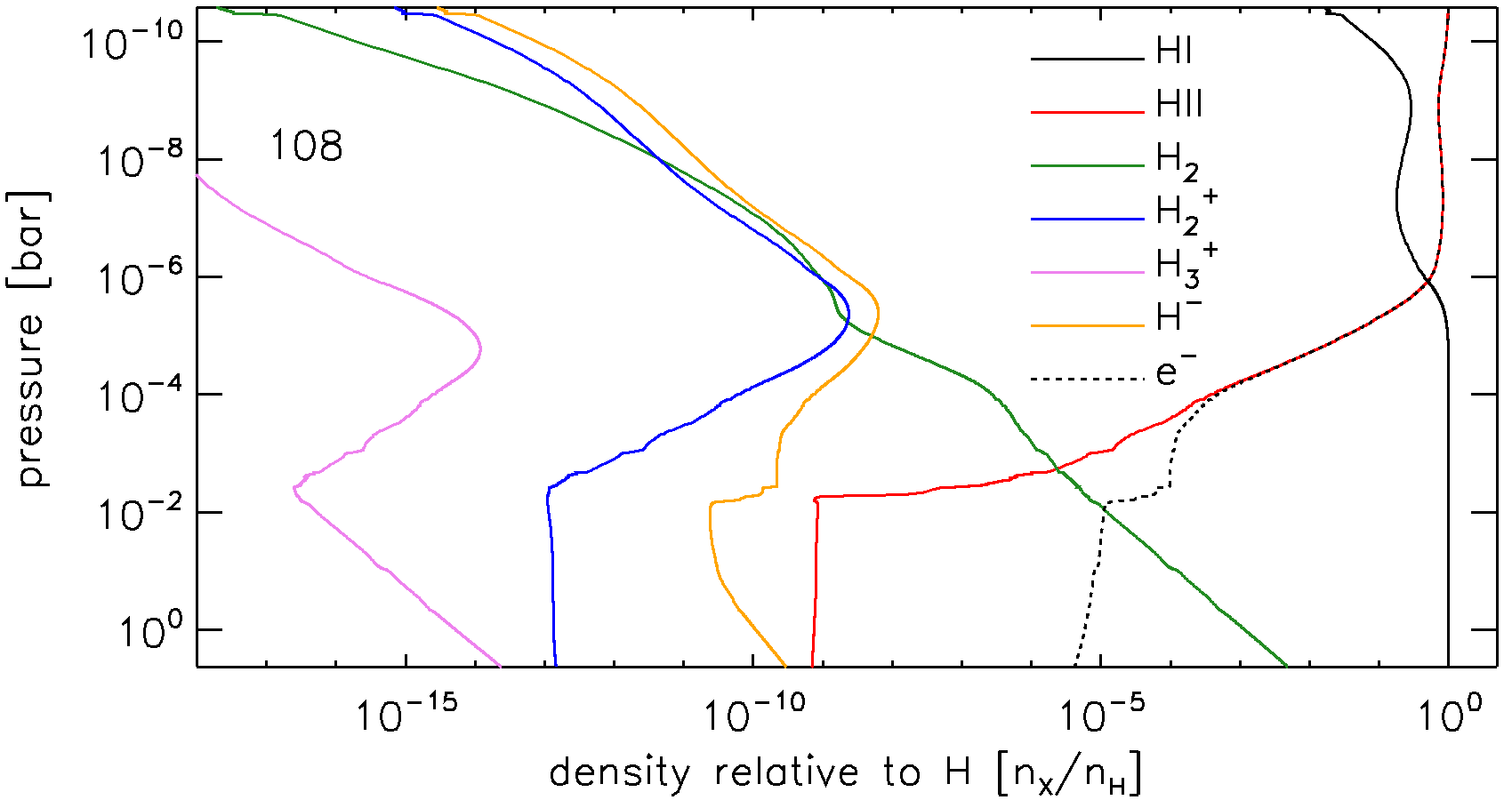}
\caption{Density relative to the total density of hydrogen for neutral hydrogen (H{\sc i}; black solid), protons (H{\sc ii}; red), molecular hydrogen (H$_2$; green), H$_2^+$ (blue), H$_3^+$ (violet), H$^-$ (orange), and electrons (e$^-$; black dashed) computed for the TP model number 025 (top) and 108 (bottom).}
\label{fig:Hcomposition}
\end{figure}

Figure~\ref{fig:Hcomposition} shows the details of the atmospheric hydrogen composition, namely the densities of neutral hydrogen (H{\sc i}), protons (H{\sc ii}), H$^-$, molecular hydrogen (H$_2$), H$_2^+$, H$_3^+$, and electrons (e$^-$) with respect to the total hydrogen density, as a function of pressure for the TP models 025 and 108. For these calculations, we employed a geometry in which the atmosphere is illuminated from the top. 

For the cooler TP profile, the atmosphere is dominated by neutral hydrogen at pressures higher than about 10$^{-9}$\,bar, while below that pressure protons are the most abundant species. For the hotter TP profile, instead, this threshold moves at a higher pressure of about 10$^{-6}$\,bar. 

At the bottom of the atmosphere, at pressures higher than about 10$^{-3}$\,bar, the atmosphere is dominated by H{\sc i} and H$_2$, with the latter decreasing rapidly with decreasing pressure. The third most abundant hydrogen species in the lower atmosphere is H$^-$, but H{\sc ii} becomes quickly more abundant, at pressures of about 10$^{-2}$\,bar and $>$1\,bar for the cooler and hotter TP models, respectively. This is mostly due to thermal ionisation, because hydrogen photoionisation occurs at higher altitudes and it is believed to be small due to the shape of the stellar spectral energy distribution \citep{fossati2018b}. Despite this, at pressures higher than about 10$^{-8}$\,bar, the profiles obtained considering the hotter TP model indicate that H$^-$ is on average about 10 times more abundant than what obtained considering the cooler TP model. This has an effect on the continuum level at optical and near infrared wavelengths that in ultra-hot Jupiters is controlled mostly by H$^-$ and H{\sc i} bound-free opacities \citep[e.g.,][]{arcangeli2018,parmentier2018}.

\begin{figure}[h!]
\includegraphics[width=\hsize,clip]{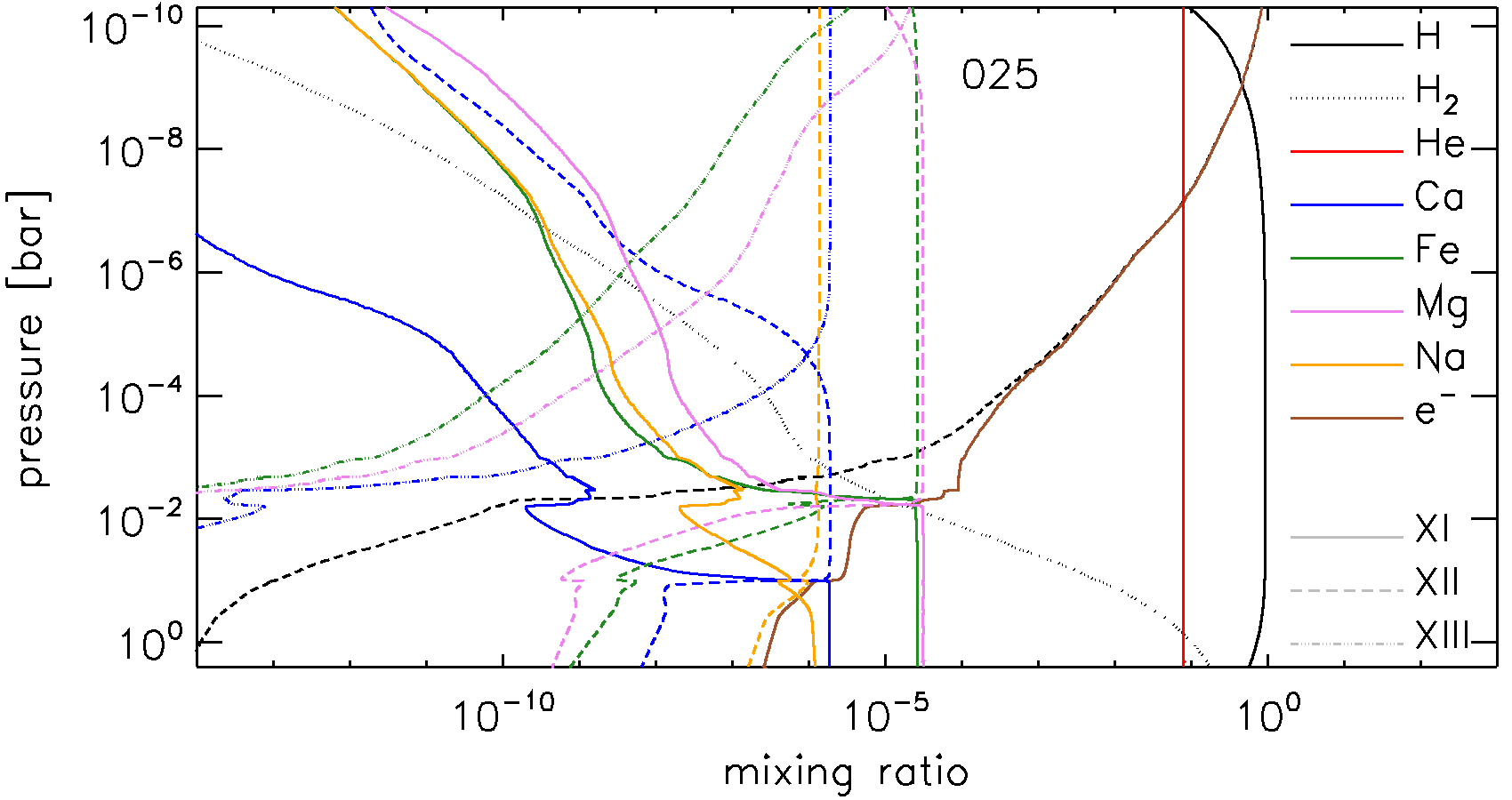}
\includegraphics[width=\hsize,clip]{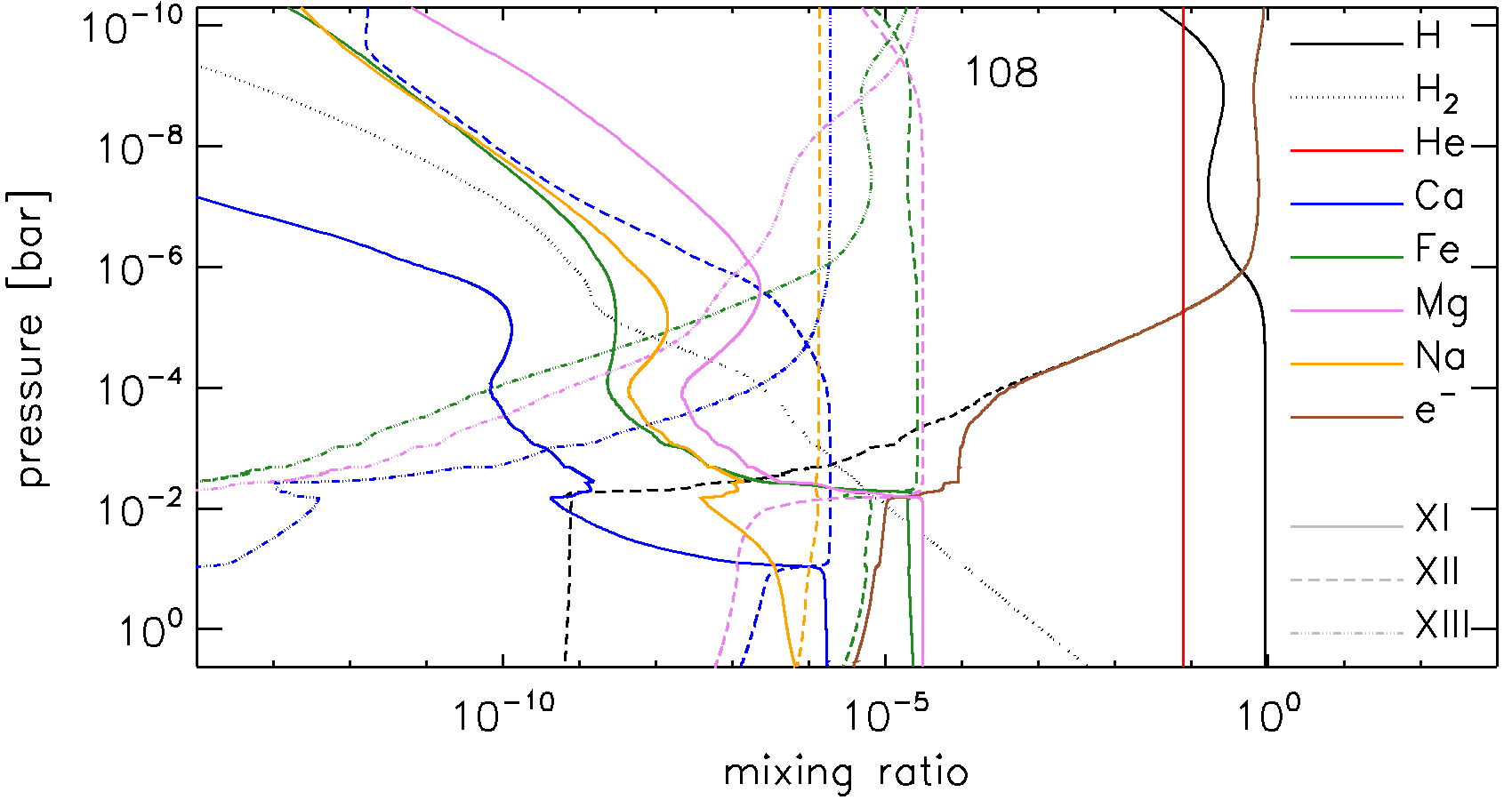}
\caption{Mixing ratios for hydrogen (black), H$_2$ (black-dotted), He (red), Ca (blue), Fe (green), Mg (violet), Na (orange), and electrons (brown) as a function of atmospheric pressure. Neutral (X{\sc i}), singly ionised (X{\sc ii}), and doubly ionised (X{\sc iii}) species are marked by solid, dashed, and dash-dotted lines, respectively. The top panel is for the TP model number 025, while the bottom panel is for the TP model number 108.}
\label{fig:all_composition}
\end{figure}

Figure~\ref{fig:all_composition} shows the mixing ratio for some of the species observationally most relevant as a function of pressure for the TP models number 025 and 108. Because of the assumption of a solar composition, He{\sc i} is the second most abundant species throughout most of the atmosphere. The distribution of Ca atoms is similar in the two cases, with Ca{\sc i} dominating below the $\sim$0.1\,bar level, Ca{\sc ii} dominating up to the $\sim$10$^{-5}$\,bar level, above which Ca{\sc iii} is the dominant Ca species. For the cooler TP model, Na is mostly neutral at pressures higher than 0.1\,bar, while it is singly ionised throughout the whole atmosphere for the hotter TP model. 

Because of their similar ionisation potentials, Mg and Fe present comparable behaviours, with Fe{\sc i} and Mg{\sc i} being the dominant species at pressures higher than 10--100\,mbar, while Fe{\sc ii} and Mg{\sc ii} are more abundant at lower pressures. Interestingly, for the hotter TP model, at pressures below 10$^{-6}$\,bar there is a mixture of almost equally abundant Fe{\sc ii} and Fe{\sc iii}. It may be therefore possible to further constrain the TP profile in KELT-9b by looking for the presence of Fe{\sc iii} lines in the planetary transmission spectrum. Some Fe{\sc iii} resonance lines are conveniently located in the near-ultraviolet (3200--3300\,\AA) and in the blue part of the optical spectrum (4000-5000\,\AA), hence where the stellar flux is high, enabling one to obtain high-quality spectra with a range of facilities such as HST, CUTE \citep{fleming2018}, and high-resolution ground-based spectrographs.
\subsection{Comparison with previous results}\label{sec:comparison_models}
We compare here Cloudy mixing ratios obtained for the cooler TP model, which is comparable to that computed with PHOENIX, with the results of \citet{kitzmann2018} and \citet{lothringer2018}. However, one has to keep in mind that our runs do not include molecules, except for H-bearing molecules, while \citet{kitzmann2018} and \citet{lothringer2018} consider a range of molecules. \citet{kitzmann2018} employed an inverted TP profile with the upper atmospheric temperature about 1500\,K cooler than that of our TP model number 025, and 1700\,K cooler than that computed with PHOENIX. Furthermore, our calculations, as well as those of \citet{kitzmann2018}, account for both thermal ionisation and photoionisation, while the results of \citet{lothringer2018} were obtained considering only thermal ionisation.

The Cloudy simulation indicates that at a pressure of about 2\,bar, our upper pressure boundary, the mixing ratios of H{\sc i} is higher than that of H$_2$ implying that the two mixing ratios become equal at even higher pressures and in particular higher than those obtained by both \citet[][$\sim$10$^{-2}$\,bar]{lothringer2018} and \citet[][$\sim$2\,bar]{kitzmann2018}. The Fe mixing ratios computed by Cloudy are very similar to those obtained by \citet{kitzmann2018}, who also found that Fe{\sc ii} becomes the dominant species at pressures lower than 10\,mbar. We obtain a good match also for the Na mixing ratio with what presented by \citet{lothringer2018}. A significant difference is found instead for the electron mixing ratio in the upper part of the atmosphere, where Cloudy gives a $\sim$100 times larger electron density than that given by \citet{lothringer2018}. We ascribe this to the fact that Cloudy considers photoionisation, while PHOENIX does not \citep{lothringer2018}. There is also a difference in the H$^-$ abundance between what we obtained with Cloudy and what presented by \citet{lothringer2018}. While their H$^-$ mixing ratio decreases almost monotonically with decreasing pressures, our H$^-$ mixing ratio increases up to 10$^{-4}$--10$^{-5}$\,bar to then decrease monotonically at lower pressures (see Fig.~\ref{fig:Hcomposition}). Since at high pressures the electron density can be considered to be consistent between the two computations, we ascribe this difference to different implementations of the physics controlling H$^-$. It will be important in the future to perform a detailed comparison, because of the importance of H$^-$ in controlling the planetary atmospheric continuum, and hence the predicted strength of spectral lines.
\subsection{Relevance of NLTE effects}\label{sec:nlte}
\begin{figure}[h!]
\includegraphics[width=\hsize,clip]{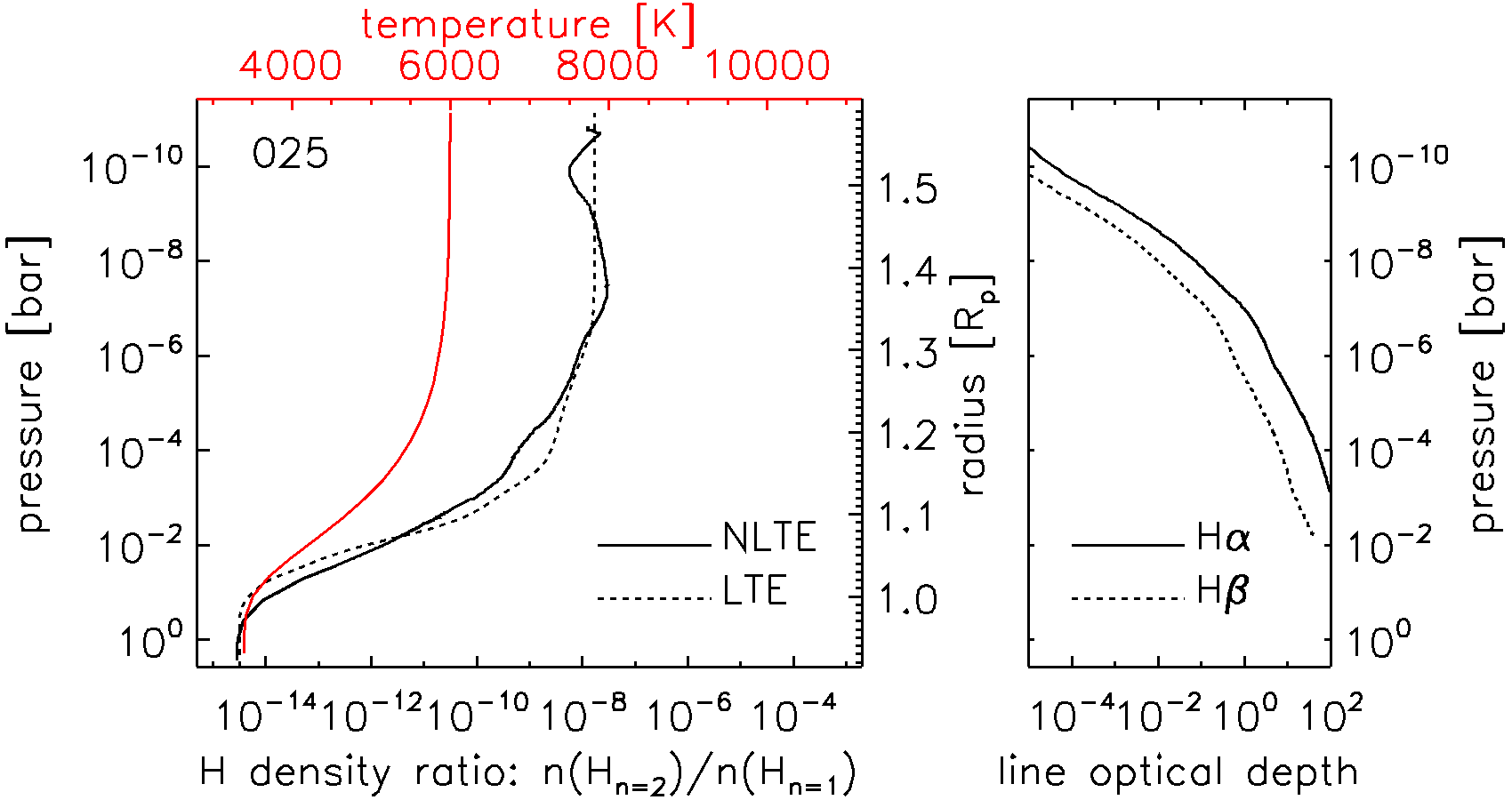}
\includegraphics[width=\hsize,clip]{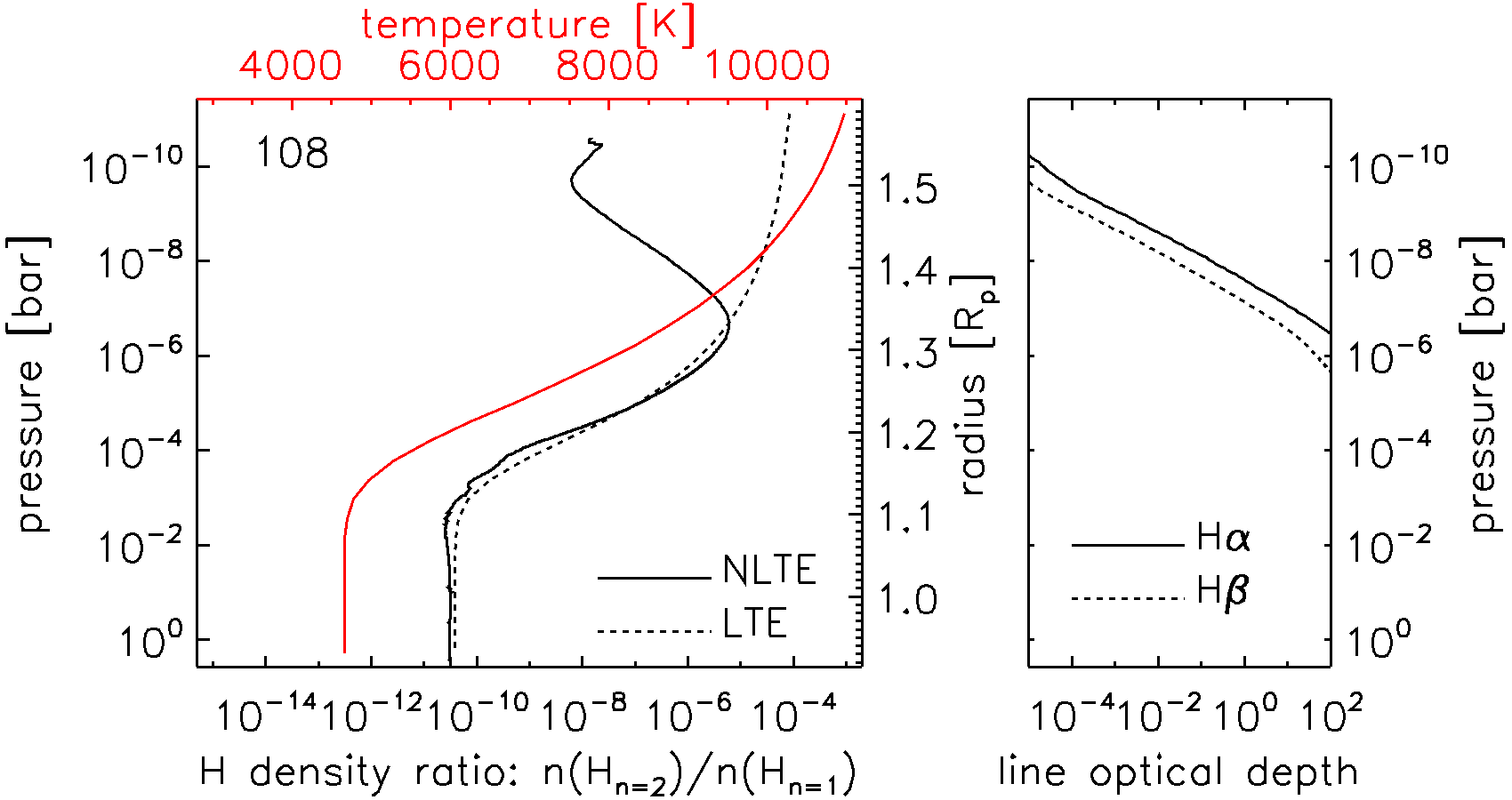}
\caption{Top-left: ratio of the n\,=\,2 and n\,=\,1 level populations for neutral hydrogen (bottom x-axis) as a function of atmospheric pressure (left y-axis) and planetary radius (right y-axis) computed with Cloudy in NLTE (solid line) and using the Boltzmann equation (i.e., LTE; dashed line) considering the TP model number 025, shown by the red line (red top x-axis). Top-right: line optical depth as a function of atmospheric pressure for H$\alpha$ (solid line) and H$\beta$ (dashed line) computed with Cloudy considering the TP model number 025. Both hydrogen lines form mostly between 10$^{-5}$ and 10$^{-8}$\,bar. Bottom: same as top panels, but for the TP model number 108.}
\label{fig:nlte}
\end{figure}

We took advantage of the NLTE radiative transfer capabilities of Cloudy to study the presence and impact of NLTE effects in the modelling of the \ha\ and \hb\ lines. To this end, we ran Cloudy for the TP models 025 and 108 employing a geometry in which the atmosphere is illuminated from the top, extracting then the NLTE hydrogen level populations and the \ha\ and \hb\ optical depth as a function of atmospheric pressure. We further used the TP models to extract the ratio between the densities of hydrogen in the n\,=\,2 and n\,=\,1 levels employing the Boltzmann equation, hence assuming LTE. Figure~\ref{fig:nlte} shows the results of this analysis.

A comparison of the top- and bottom-left panels of Fig.~\ref{fig:nlte} indicates that the hotter TP model indeed leads to a larger density of excited hydrogen atoms by a factor of 10$^3$--10$^4$ compared to what obtained from the cooler TP model, hence stronger Balmer lines better fitting the observations. Figure~\ref{fig:nlte} also shows that, except for the case of the hotter TP profile and at pressures below $\sim$10$^{-6}$\,bar, there seems to be a general good match between the LTE and NLTE n\,=\,2 and n\,=\,1 hydrogen level populations, possibly indicating that the LTE approximation might be valid for computing the hydrogen Balmer lines. However, the profiles obtained from TP model number 025 indicate that at pressures corresponding to the main \ha\ and \hb\ line formation region (i.e., between 10$^{-5}$ and 10$^{-8}$\,bar; right) the n\,=\,2 LTE level population is almost 10 times smaller than that computed accounting for NLTE effects, hence leading to LTE underestimating the strength of the lines. For the hotter TP model, the LTE and NLTE level populations at pressures below $\sim$10$^{-6}$\,bar diverge significantly, leading to LTE overestimating the strength of the lines. We also note that, considering an isothermic profile with a temperature similar to that of the upper atmosphere of TP model number 108, \citet{wyttenbach2020} found an even larger difference (about 10$^6$) in the LTE population of excited hydrogen compared to NLTE, but their result is based on constant temperature and density ratio profiles that are further strongly correlated with the mass-loss rate, which hamper comparisons between the two results.

By construction, Cloudy is unable to perform calculations of line profiles assuming LTE for the n\,=\,1 and n\,=\,2 hydrogen levels. For this reason, we computed the \ha\ transmission line profiles assuming LTE approximation as follows. First, we computed the LTE number densities of excited (i.e., n\,=\,2) hydrogen using the Boltzmann formula, further considering the lowering of the ionisation potential due to Debye screening by electrons and ions in the calculation of the partition function. Next, we assumed the absorption profile to be described by a Voigt function and considered temperature (Doppler), natural, and pressure (Stark and van der Waals) broadening. Because of the relatively low gas density in the region of the atmosphere probed by transmission spectroscopy, pressure broadening is weak, in agreement with our Cloudy simulations and the results of \citet{wyttenbach2020}. We additionally cross checked our calculations by using a more elaborate treatment of the \ha\ profile employing pressure broadening tables after \citet{lemke1997} obtaining a good match between the two approaches. This confirms that a Voigt profile is a good approximation for interpreting transmission observations of the \ha\ line. All relevant numerical routines were taken from the \textsc{LLmodels} stellar model atmosphere code \citep{shulyak2004}. To calculate the LTE transmission profiles we employed the $\tau$-REx (Tau Retrieval for Exoplanets) forward model \citep{waldmann2015a}. We used an extended version of the original package, which now includes additional continuum opacity sources relevant for the high atmospheric temperatures of KELT-9b, as described in \citet{shulyak2020}.

Because \textsc{LLmodels} is a module structured code, we turned some of its modules into numerical libraries and loaded them into $\tau$-REx to compute continuum opacity additional to those described in \citet{shulyak2020}. In particular, we calculated opacities due to H{\sc i}, He, and various metals, which contribute a small, but non-negligible, $\sim$8\% to the total transmission depth at the wavelengths around the \ha\ line.

The number densities of neutral and ionised species were taken from the Cloudy calculations for the respective model structures. This ensures that the background opacity is as similar as possible between the $\tau$-REx and Cloudy computations, enabling us to isolate and study the effect of NLTE on the \ha\ transmission spectra.

\begin{figure}[h!]
\includegraphics[width=\hsize,clip]{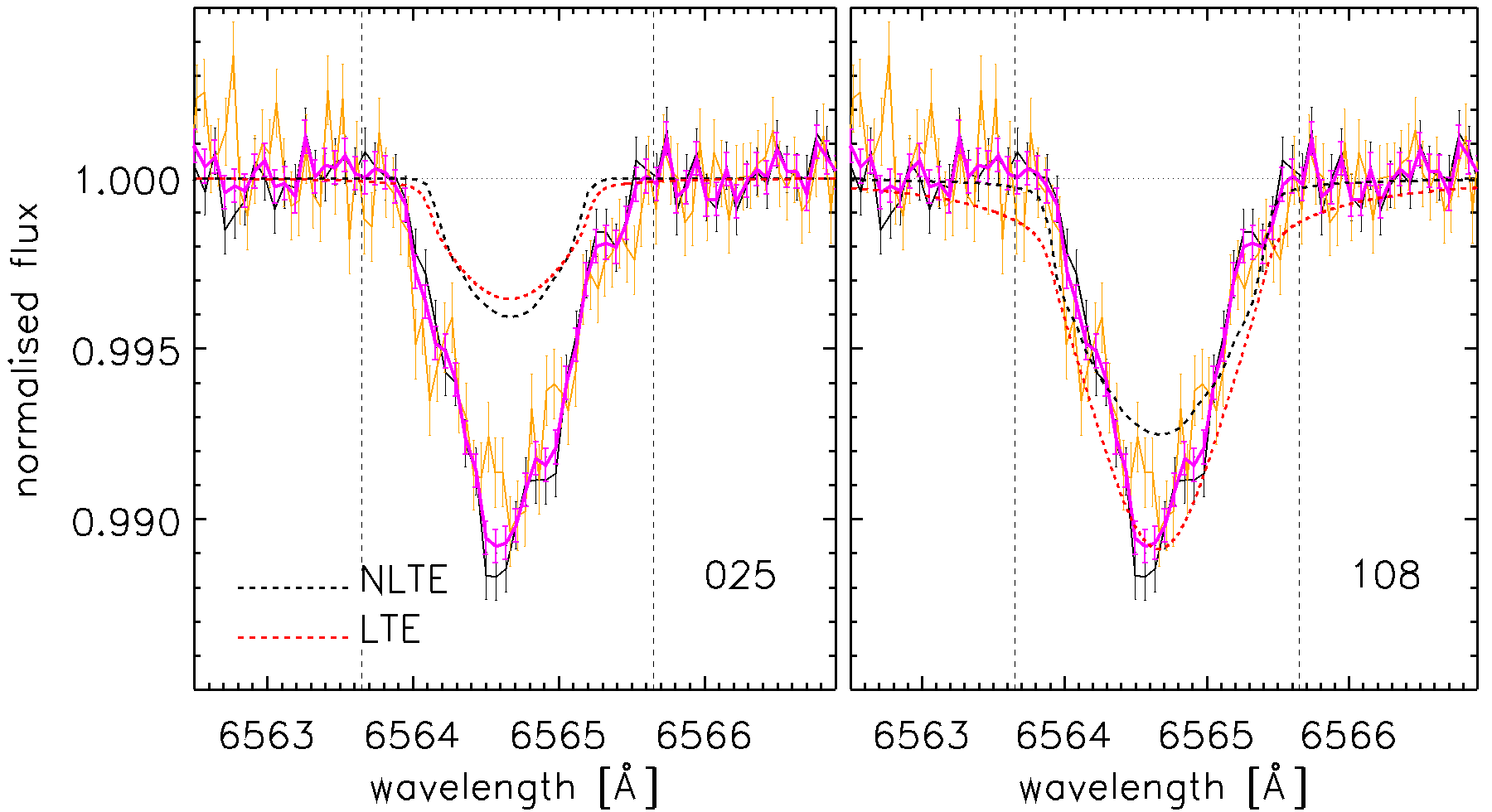}
\caption{Comparison between observed and synthetic \ha\ line profiles computed in LTE (red dashed line) and NLTE (black dashed line) considering the TP model numbers 025 (left) and 108 (right). The colour scheme of the observed line profiles is the same as in Fig.~\ref{fig:comparison_observations}.}
\label{fig:nlte_Halpha}
\end{figure}

Figure~\ref{fig:nlte_Halpha} presents the comparison of the LTE and NLTE \ha\ transmission spectra computed considering the TP model number 025 and 108. As expected from the comparison of the level populations, the LTE \ha\ profiles obtained considering the TP model number 025 is comparable to that obtained considering NLTE effects. Instead, with the hotter model, the LTE \ha\ transmission spectrum is stronger than the one obtained considering NLTE, particularly for the line wings, though the line core appears to be a good fit, in a way similar to our model producing the strongest lines (TP model number 114; see Figure~\ref{fig:profilesANDtp}). This difference is caused by the difference in the LTE and NLTE level populations at pressures lower than 10$^{-6}$\,bar.   

In agreement with \citet{garcia2019}, our simulations indicate that NLTE effects are extremely important for correctly modelling the hydrogen Balmer lines in the transmission spectrum of KELT-9b. Although \citet{wyttenbach2020} do not find strong evidence in favor of NLTE, they find a departure from LTE of the order of 10$^6$ and assumed a constant density ratio of excited hydrogen to total hydrogen throughout the atmosphere, which our simulations clearly indicate being not realistic, particularly for the hotter models \citep[i.e., temperatures similar to those obtained by][]{wyttenbach2020}. Indeed, contrary to \citet{wyttenbach2020}, who considered only a simple scaling for the hydrogen n\,=\,2 level population, we carried out realistic and self-consistent NLTE calculations taking into account the relevant radiative and collisional transitions between bound and free states in the hydrogen atom.

In an attempt to more meaningfully compare our LTE calculations with those of \citet{wyttenbach2020}, we computed an LTE \ha\ profile assuming an isothermal temperature at T\,=\,13000\,K, which is the LTE temperature profile obtained by \citet{wyttenbach2020} best fitting the observed \ha\ line, and the chemical abundances obtained from a Cloudy run employing the same isothermal profile. However, we find that the depth of our LTE \ha\ profile constructed under these assumptions is about twice that of \citet{wyttenbach2020} and that it would require to decrease the density of excited hydrogen by a factor of ten to be able to fit the observations. It is possible that this further discrepancy is caused by differences in the underlying chemical composition, i.e. Cloudy vs equilibrium chemistry without photoionisation. It is also possible that meaningful comparisons of our results with those of \citet{wyttenbach2020} are in general not possible, because they considered the planetary mass-loss rate as a free parameter, further obtaining that it strongly correlates with the atmospheric temperature and hydrogen density profiles.
\subsection{Impact of stellar effective temperature}\label{sec:teff}
So far, we attempted to fit the observed hydrogen Balmer lines by modifying the TP profile, obtaining that it should be significantly different than that obtained self-consistently with PHOENIX. However, there are other ways of modifying the line profiles without necessarily changing significantly the TP profile, for example by decreasing the mean molecular weight, hence increasing the pressure scale height, and/or by increasing the temperature of the host star, hence increasing the atmospheric energy available for exciting hydrogen. Therefore, we discuss whether increasing the effective temperature of the planetary atmosphere (this section) or decreasing the metallicity of the planetary atmosphere (next section) would enable one to fit the observed hydrogen Balmer lines employing the self-consistently computed TP profile. Furthermore, looking for the possible effect of a different stellar spectral energy distribution is important at the light of the gravity darkening effect recently detected for KELT-9, with the temperature ranging from 10200\,K at the poles and 9400\,K at the equator, and with the stellar inclination angle tilted by almost 40 degrees \citep{ahlers2020}.

As shown by \citet{lothringer2019}, the shape of the input stellar spectral energy distribution has an impact on the atmospheric properties of ultra-hot Jupiters \citep[see also][]{fossati2018b}. Therefore, we checked whether a variation in the temperature of the star compatible with what given by stellar spectroscopic analyses leads to significant changes in the strength of the \ha\ and \hb\ line profiles.
\begin{figure}[h!]
\vspace{0.5cm}
\includegraphics[width=\hsize,clip]{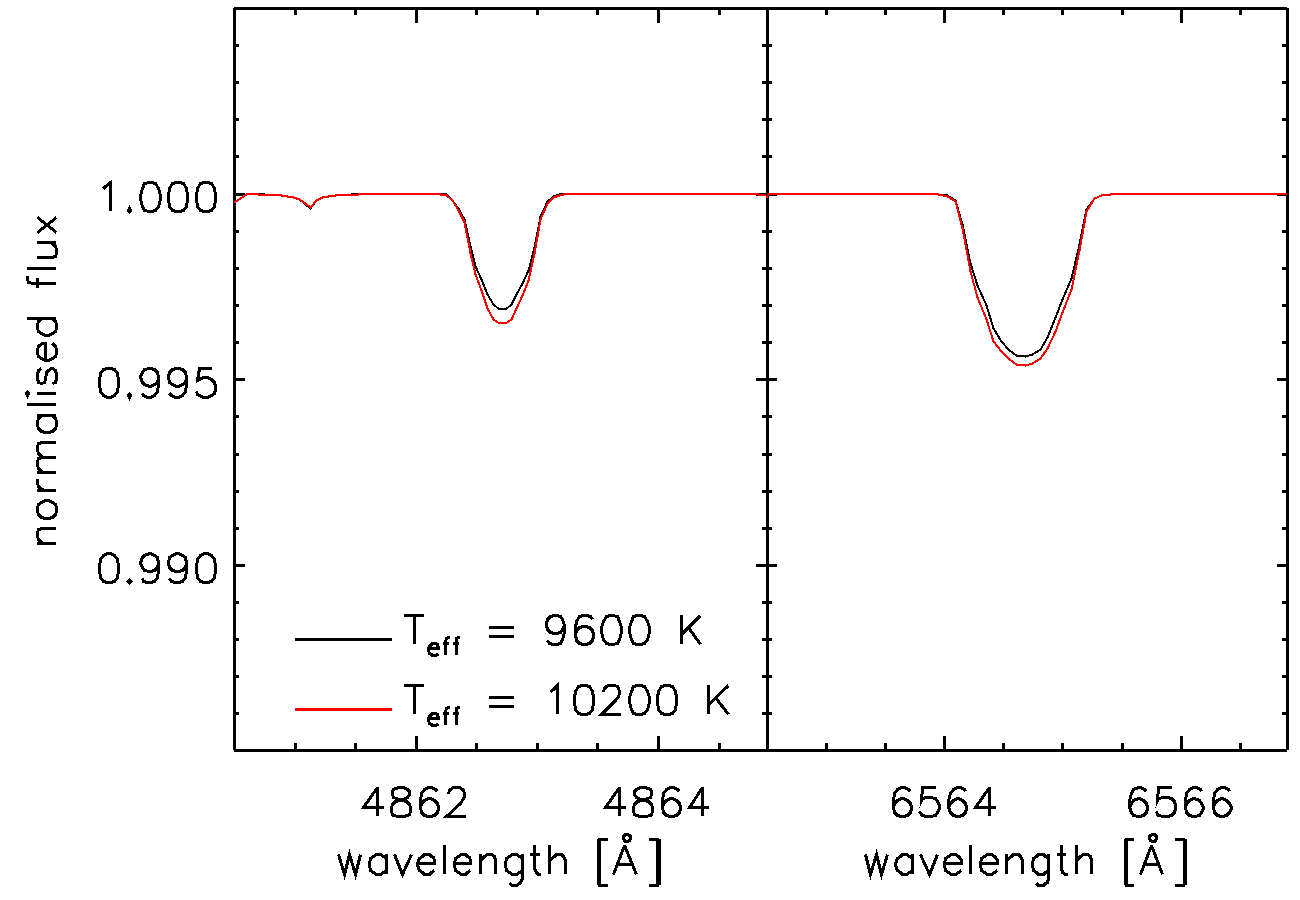}
\caption{Comparison between the H$\beta$ (left) and H$\alpha$ (right) line profiles computed in NLTE with Cloudy employing an incident stellar flux corresponding to a star with an effective temperature of 9600\,K \citep[black;][]{borsa2019} and 10200\,K \citep[red;][]{gaudi2017}. To ease the comparison, the range of the y-axis is the same as that of the top panels of Fig.~\ref{fig:profilesANDtp}.}
\label{fig:teff}
\end{figure}

To this end, we calculated a further transmission spectrum with Cloudy considering the TP model number 025 and a stellar spectral energy distribution computed with an effective temperature of 10200\,K, hence 600\,K hotter than what employed for the analysis presented in Sect.~\ref{sec:results}. We remark that the stellar effective temperature we adopted in this work is that given by \citet{borsa2019}, while the hotter one we chose for studying the impact of stellar effective temperature on the transmission spectrum is that given by \citet{gaudi2017}.

Figure~\ref{fig:teff} shows the comparison between the \ha\ and \hb\ line profiles obtained for the two stellar spectral energy distributions and considering the TP model number 025. The hotter star leads to a slight increase in the level population of excited hydrogen and therefore to stronger lines, but the increase is small compared to what would be necessary to fit the observed profiles. This indicates that fitting the observed line profiles requires a planetary atmosphere with a temperature as high as $\sim$10000--11000\,K, even when considering a hotter star. This result further indicates that the fact that the planet crosses regions of the star with different temperatures due to the fast stellar rotation \citep{ahlers2020} has no detectable effect on the transmission spectra.
\subsection{Planetary atmospheric metallicity}\label{sec:metallicity}
Following Eq.~(\ref{eq.scale_height}), it is possible to increase the pressure scale height and hence the size of the absorption features, without exceedingly increasing the temperature, by decreasing the mean molecular weight. Therefore, we computed a synthetic transmission spectrum covering the \ha\ and \hb\ line profiles, employing the TP profile number 025 and a metallicity of 0.1$\times$solar. 
\begin{figure}[h!]
\includegraphics[width=\hsize,clip]{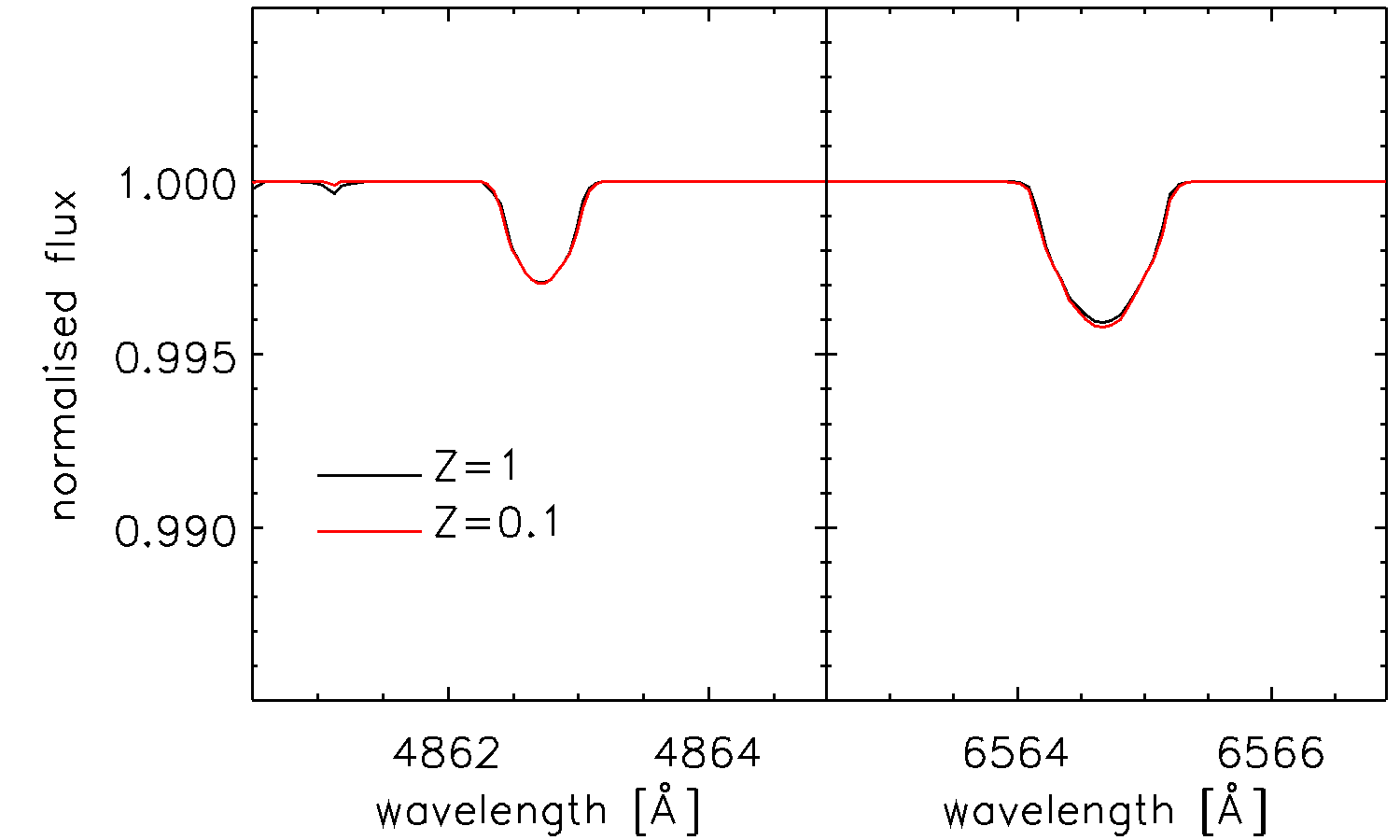}
\caption{Comparison between the H$\beta$ (left panel) and H$\alpha$ (right panel) transmission profiles obtained considering the TP profile number 025 and a solar metallicity (Z\,=\,1; black line) or a sub-solar metallicity (Z\,=\,0.1; red line). To ease the comparison, the range of the y-axis is the same as that of the top panels of Fig.~\ref{fig:profilesANDtp}.}
\label{fig:comparison_Vturb0_lowZ}
\end{figure}

Figure~\ref{fig:comparison_Vturb0_lowZ} shows the comparison between the \ha\ and \hb\ line profiles obtained for the TP model number 025 and considering a solar-like and sub-solar planetary atmospheric metallicity. Decreasing the metallicity only very slightly increased the depth of the absorption lines further confirming that fitting the observations requires a planetary atmosphere with a temperature of $\sim$10000--11000\,K.
\subsection{Possible origin and consequences of the additional atmospheric heating}\label{sec:origin}
Our results indicate that the atmosphere of KELT-9b, particularly at pressures lower than $\sim$10$^{-3}$\,bar, should be significantly hotter than that estimated by self-consistent calculations, e.g. with PHOENIX. However, the difficulty of simultaneously fitting the strength and the width of the \ha\ and \hb\ line profiles suggests that either our results slightly overestimate the upper atmospheric temperature or the analysis of the observations, possibly the spectral normalisation, partially removes the outer edges of the line wings. We find that the upper atmospheric temperature of the family of TP profiles leading to best fit the observations is around 10000--11000\,K that is comparable to that of the stellar photosphere, which is the main heating source. This temperature also reminds of the upper atmospheric temperature of classical hot Jupiters orbiting late-type stars \citep[e.g.,][]{yelle2004}. However, there are significant differences in the heating source and mechanisms between classical hot Jupiters and ultra-hot Jupiters like KELT-9b orbiting early A-type stars \citep{fossati2018b}.

The upper atmospheres of planets orbiting stars later than spectral type A3-A4 are heated by absorption of the stellar high-energy emission, namely X-ray and EUV (XUV) radiation. The temperature reached in the upper atmosphere of these planets is typically of the order of 10$^4$\,K, which is also of the same order of magnitude as that of the plasma in the stellar chromosphere and transition region emitting the majority of the XUV radiation absorbed by the planet and causing the heating. Planets orbiting early A-type stars, instead, are believed not to possess a chromosphere and transition region, hence not to emit significant XUV radiation\footnote{Early A-type stars emit some EUV radiation, but it is of photospheric origin and too weak and at too long wavelengths to produce significant heating in planetary upper atmospheres \citep{fossati2018b}.} \citep{fossati2018b}. Indeed, the atmospheres of ultra-hot Jupiters are heated primarily by the less energetic UV and optical stellar photons. Given that the stellar photospheric temperature is of the order of 10000\,K, from a thermodynamic point of view it would be possible for the planetary atmosphere to reach such values. For the same reason, it is then unlikely for the planetary atmosphere to become much hotter, unless there is significant atmospheric heating due to other mechanisms, such as ohmic dissipation or magnetohydrodynamic waves \citep[e.g.,][]{thorngren2018}. This conclusion is also supported by detailed hydrodynamic simulations of the upper atmosphere of KELT-9b conducted by \citet{mitani2020}.

It is, however, interesting to notice that the upper atmospheric temperature of the TP profile we consider as reference for leading to best fit the observations lies at the upper limit of the explored temperature range. Therefore, we generated 30 more TP profiles with a lower atmospheric temperature close to that of the TP profile number 108 and upper atmospheric temperatures ranging between 11000 and 16000\,K. We further produced synthetic transmission spectra on the basis of these profiles and compared them with the observations in the same way as described in Sect.~\ref{sec:transmission} and Sect.~\ref{sec:results}. We obtained that four of these hotter models satisfied the conditions set in Eq.~(\ref{eq:conditions}) and that three models satisfied the more restrictive conditions set in Eq.~(\ref{eq:conditions_strict}). However, all these models display an upper atmospheric temperature in excess of 12000\,K, which, given the stellar effective temperature, we consider not being realistic for describing the TP profile of KELT-9b. Furthermore, none of these models leads to a better fit to the data with respect to the TP profile number 108 and also have the problem of being shallower and broader than the observations.

It is therefore possible that self-consistent models computing the planetary TP structure underestimate atmospheric heating and/or overestimate cooling. \citet{garcia2019} showed that, because of the shape of the stellar spectral energy distribution and of the large density of excited hydrogen in the planetary atmosphere, heating through absorption of the stellar UV and optical radiation plays an important role.

In Sect.~\ref{sec:Hcomposition}, we showed that the planetary atmospheric composition we obtained from Cloudy differs from those of \citet{kitzmann2018} and \citet{lothringer2018} also in the electron density, which we find being significantly higher, particularly in the upper atmosphere. Indeed, the Cloudy runs suggest a higher density of ionised metals and a higher metal ionisation degree compared to previous models, particularly in the upper atmospheric layers. This may be due to the fact that, in contrast to PHOENIX, Cloudy includes metal photoionisation. Given that metal absorption and ionisation are among the main heating mechanisms in ultra-hot Jupiters, it is possible that self-consistent calculations of the atmospheric TP profile accounting for photoionisation would lead to metal ionisation properties similar to those found with Cloudy, hence to increased upper atmospheric temperatures.

An upper atmosphere at a temperature of 10000--11000\,K would also explain the strong Cr{\sc ii} and Fe{\sc ii} absorption features obtained by \citet{hoeijmakers2019} from the analysis of optical high-resolution transmission spectroscopy observations. The observed features were surprisingly large compared to the model, which was produced considering an isothermal temperature profile at 4000\,K, while a higher atmospheric temperature leads to a stronger Fe and Cr ionisation, hence to larger features of both species. Therefore, it would be important to re-analyse these observations considering a model computed on the basis of a hotter TP profile (e.g., the TP model number 108), which would lead to increase the detection level of several ions and possibly also to the detection of additional elements that would constrain the shape of the atmospheric TP profile (e.g., Fe{\sc iii}).
\subsection{Impact of the model assumptions}\label{sec:assumptions}
By employing Cloudy, we were able to lift the LTE approximation for modelling lines rising from excited hydrogen levels and therefore run all computations accounting for NLTE effects for all considered elements. Despite this, the synthetic hydrogen Balmer line profiles did not fit well enough the observations and therefore we discuss here the modelling assumptions that may possibly play a role in this outcome, focusing in particular on the assumption of a 1D geometry and of a hydrostatic atmosphere.

Because of the 1D approximation, we assumed that the day and night side of the planetary atmosphere through which the stellar light travels have the same properties. This is clearly not the case, though global circulation models of ultra-hot Jupiters suggest that there is a relatively small temperature (i.e., 1000--1500\,K) and compositional difference between the day and the night side \citep[e.g.,][]{bell2018,tan2019} and phase curve observations of KELT-9b confirm this prediction \citep{wong2019,mansfield2020}, hence alleviating the effect of this assumption on our results. The 1D assumption likely leads to overestimate the number of excited hydrogen atoms, but also to overestimate hydrogen ionisation. These two effects counteract each other in terms of their impact on the strength of the hydrogen Balmer lines and future simulations accounting for the three-dimensional nature of the planetary atmosphere are therefore necessary to assess how the 1D assumption affects synthetic transmission spectra of KELT-9b.

The assumption of a hydrostatic atmosphere may have two important effects on our results: 1) the atmospheric layers characterised by strong upward acceleration (i.e., in a hydrodynamic state) have a larger pressure scale height, because the upward acceleration reduces locally the planetary gravity; 2) a hydrodynamic atmosphere is also characterised by a significant upward transport of gas from the bottom layers, which would for example lead to increase the density of (excited) neutral hydrogen atoms higher up in the atmosphere. Both these  effects would lead to an increase in the strength of the \ha\ and \hb\ lines, without the need of increasing the atmospheric temperature. \citet{fossati2018b}, \citet{garcia2019}, and \citet{mitani2020} showed that the upper atmosphere of KELT-9b is probably expanding hydrodynamically.

Within Cloudy, to be able to fix the temperature profile, we had to exclude from the calculations all non-hydrogen-based molecules, which may have affected the lower atmosphere. However, our results indicate that the lower atmosphere has a negligible impact on the shape and strength of the hydrogen Balmer lines. Furthermore, the overall hotter atmospheric temperature would lead to an even lower molecular abundance, decreasing the impact these would have on the results.
\section{Conclusion}\label{sec:conclusion}
We constructed 126 empirical atmospheric temperature-pressure profiles for the ultra-hot Jupiter KELT-9b varying the lower and upper atmospheric temperatures, and the location and gradient of the temperature rise, further considering the available observational constraints \citep{wong2019,mansfield2020,pino2020,turner2020}. We then employed the Cloudy NLTE radiative transfer code in a one-dimensional geometry to produce transmission spectra of the \ha\ and \hb\ lines on the basis of the constructed TP profiles, comparing them with the available observations.

We found that the family of TP profiles leading to best fit the observations is characterised by an upper atmospheric temperature of 10000--11000\,K, hence about 4000\,K hotter than predicted by the PHOENIX model, and by a temperature rise starting at pressures higher than 10$^{-4}$\,bar, which is around the highest pressure level probed by the \ha\ and \hb\ lines. For the TP profile leading to best fit the observations, we compared the n\,=\,2 excited hydrogen level population and the \ha\ and \hb\ line profiles computed in LTE and NLTE obtaining that the LTE approximation leads to overestimate the level population by several orders of magnitude, and hence to overestimate the strength of the lines. Furthermore, the Cloudy simulations also clearly indicate that other commonly considered assumptions, such as constant temperature and/or density ratio profiles, are not realistic and would lead to misinterpret the data.

We further analysed the atmospheric chemical composition obtained with Cloudy employing the TP profile leading to best fit the \ha\ and \hb\ lines. The main results of this analysis are that the high upper atmospheric temperature is likely caused by metal photoionisation and that at pressures lower than about 10$^{-6}$\,bar, Fe{\sc iii} is almost as abundant as Fe{\sc ii}. The latter finding indicates that Fe{\sc iii} may be also detectable in the transmission spectra of KELT-9b. Its identification in the observed transmission spectra would be a confirmation that the upper atmospheric temperature of KELT-9b is significantly higher than previously thought. More in general, the detection and modelling of features rising from different ions of the same element and from different electronic levels of the same ion is important to constrain the physical properties of the planetary atmosphere, including the local thermodynamic state, which is a key piece of information for the further development of planetary atmosphere models. However, the detection of such metal features should be based on synthetic spectra computed with adequate TP profiles and accounting for NLTE effects that become relevant when looking at lines rising from excited states, as is the case of the hydrogen Balmer lines. 

The family of TP profiles leading to best fit the observed hydrogen Balmer lines, however, leads still to profiles that are shallower and broader than the observations, though the line strength (i.e., equivalent width) is comparable to the measured one. There are a range of possible explanations for this result connected to both data analysis and modelling assumptions. The observed transmission spectra covering the hydrogen Balmer lines of KELT-9b have been obtained with different instruments and employing slightly different data analysis methods, which may have introduced systematic differences among the results. It is also not possible to exclude that the data analysis procedures have removed part of the far line wings, leading to narrower spectral lines. Therefore, we suggest that a uniform analysis of all available spectroscopic data covering the primary transit of KELT-9b would be beneficial for future theoretical studies aiming at fitting the observations.

On the modelling side, having lifted the LTE approximation, the two main assumptions affecting the results are those of a 1D geometry and of a hydrostatic atmospheres. The impact of the former on the synthetic spectra is unclear, because of the competing effects (lower hydrogen ionisation and lower hydrogen excitation) that the cooler nightside temperature would have on the transmission spectra. Instead, the implementation of a model accounting for hydrodynamic effects may change the shape of the line profiles precisely in the desired direction, namely making them deeper without the need of further increasing the temperature, which would further broaden the line wings. Therefore, we suggest that future modelling and fitting of the hydrogen Balmer lines of KELT-9b shall attempt to account for hydrodynamic effects in the upper atmosphere, further to accounting for NLTE effects.

Therefore, future observational works should primarily attempt to understand the origin of the differences detected in the currently available observations, particularly by collecting higher-quality data, if possible, and by analysing them in a homogeneous way. However, unfortunately, given the significant computing time currently necessary for producing synthetic spectra accounting for NLTE effects, at least for the near future, fitting the observations will be possible only through forward modelling.

An upper atmospheric temperature of 10000--11000\,K has strong implications for the planetary atmospheric mass loss. Indeed, the extra energy available as a result of such a high atmospheric temperature would lead to a significantly higher mass-loss rate compared to the 10$^{11}$\,g\,s$^{-1}$ obtained on the basis of the PHOENIX temperature-pressure profile \citep{fossati2018b}. A significantly higher mass-loss rate may have important consequences for the atmospheric evolution of the planet, but may also indicate that the mass-loss rates expected also for other ultra-hot Jupiters, particularly those orbiting hotter stars, might need to be reconsidered. Indeed, the analyses of \citet{yan2018} and \citet{wyttenbach2020} suggested for KELT-9b mass-loss rates as high as 10$^{12}$\,g\,s$^{-1}$. Interestingly, the ultra-hot Jupiters detected so far appear to be on average more massive than the classical hot Jupiters of the same radius, which is not a bias effect because all these planets have been detected with the transit method and have similar radii. Therefore, planets with masses smaller than those typical of ultra-hot Jupiters may be difficult to detect because of their very small radius as a consequence of the strong atmospheric escape. Future work should therefore also aim to constrain and estimate planetary mass-loss rates for ultra-hot Jupiters as a function of planetary parameters, to identify whether there could be a signature of mass loss in the observed planet population.
\begin{acknowledgements}
M.E.Y. acknowledges funding from the \"OAW-Innovationsfonds IF\_2017\_03. D.S. and M.R. acknowledge the support of the DFG priority program SPP-1992 ``Exploring the Diversity of Extrasolar Planets'' (DFG PR 36 24602/41). A.G.S. and L.F. acknowledge financial support from the Austrian Forschungsf\"orderungsgesellschaft FFG project CONTROL P865968. L.M.L. acknowledges the financial support from the State Agency for Research of the Spanish MCI through the ``Center of Excellence Severo Ochoa'' award SEV-2017-0709, and from the research project PGC2018-099425-B-I00. A.W. acknowledges the financial support of the SNSF by grant number P400P2\_186765. The authors thank the anonymous referee for the insightful comments.
\end{acknowledgements}
\begin{appendix}
\section{Empirical TP profiles and their parameters.}
%
\begin{table*}[]
\caption{List of empirical TP profiles and their parameters. The column labelled as $p_{0}$ gives the reference pressure, in mbar, obtained through an iterative process for each TP model (see Sect~\ref{sec:transmission}).}
\label{tab.TPprofiles}
\begin{footnotesize}
\begin{center}
\begin{tabular}{l|cccc|c||l|cccc|c||l|cccc|c}
\hline
\hline
Model \# & $\kappa$ & $\gamma$ & $\beta$ & $s$ & $p_{0}$ & Model \# & $\kappa$ & $\gamma$ & $\beta$ & $s$ & $p_{0}$ & Model \# & $\kappa$ & $\gamma$ & $\beta$ & $s$ & $p_{0}$ \\
\hline
001 & 2.5 & 1.01 & 1.02 & 1.0 & 3    & 043 & 2.5 & 0.87 & 1.10 & 2.5 & 18   & 085 & 2.5 & 0.63 & 1.23 & 4.0 & 18   \\
002 & 2.5 & 1.26 & 1.04 & 1.0 & 0.2  & 044 & 2.5 & 1.07 & 1.15 & 2.5 & 0.4  & 086 & 2.5 & 0.89 & 1.27 & 4.0 & 1    \\
003 & 2.5 & 1.50 & 1.04 & 1.0 & 0.06 & 045 & 2.5 & 1.23 & 1.21 & 2.5 & 0.06 & 087 & 2.5 & 1.06 & 1.33 & 4.0 & 0.2  \\
004 & 2.5 & 1.71 & 1.04 & 1.0 & 0.03 & 046 & 2.5 & 1.43 & 1.22 & 2.5 & 0.03 & 088 & 2.5 & 1.20 & 1.39 & 4.0 & 0.06 \\
005 & 2.5 & 1.90 & 1.04 & 1.0 & 0.03 & 047 & 2.5 & 1.60 & 1.23 & 2.5 & 0.03 & 089 & 2.5 & 1.36 & 1.42 & 4.0 & 0.03 \\
006 & 2.5 & 2.07 & 1.04 & 1.0 & 0.03 & 048 & 2.5 & 1.77 & 1.23 & 2.5 & 0.03 & 090 & 2.5 & 1.51 & 1.44 & 4.0 & 0.03 \\
007 & 2.5 & 1.09 & 0.98 & 2.5 & 18   & 049 & 2.5 & 0.89 & 1.09 & 4.0 & 18   & 091 & 3.1 & 0.42 & 1.35 & 1.0 & 3    \\
008 & 2.5 & 1.30 & 1.02 & 2.5 & 1    & 050 & 2.5 & 1.10 & 1.14 & 4.0 & 18   & 092 & 3.1 & 0.70 & 1.39 & 1.0 & 0.2  \\
009 & 2.5 & 1.51 & 1.04 & 2.5 & 0.2  & 051 & 2.5 & 1.27 & 1.19 & 4.0 & 0.4  & 093 & 3.1 & 0.93 & 1.42 & 1.0 & 0.06 \\
010 & 2.5 & 1.71 & 1.04 & 2.5 & 0.06 & 052 & 2.5 & 1.45 & 1.22 & 4.0 & 0.2  & 094 & 3.1 & 1.14 & 1.43 & 1.0 & 0.03 \\
011 & 2.5 & 1.90 & 1.04 & 2.5 & 0.03 & 053 & 2.5 & 1.64 & 1.22 & 4.0 & 0.2  & 095 & 3.1 & 1.32 & 1.44 & 1.0 & 0.03 \\
012 & 2.5 & 2.07 & 1.04 & 2.5 & 0.03 & 054 & 2.5 & 1.82 & 1.22 & 4.0 & 0.2  & 096 & 3.1 & 1.49 & 1.44 & 1.0 & 0.03 \\
013 & 2.5 & 1.09 & 0.98 & 4.0 & 47   & 055 & 3.1 & 0.70 & 1.19 & 1.0 & 3    & 097 & 3.1 & 0.44 & 1.34 & 2.5 & 18   \\
014 & 2.5 & 1.31 & 1.02 & 4.0 & 47   & 056 & 3.1 & 0.98 & 1.21 & 1.0 & 0.2  & 098 & 3.1 & 0.70 & 1.39 & 2.5 & 0.4  \\
015 & 2.5 & 1.52 & 1.04 & 4.0 & 47   & 057 & 3.1 & 1.22 & 1.22 & 1.0 & 0.06 & 099 & 3.1 & 0.93 & 1.42 & 2.5 & 0.2  \\
016 & 2.5 & 1.74 & 1.04 & 4.0 & 47   & 058 & 3.1 & 1.43 & 1.22 & 1.0 & 0.03 & 100 & 3.1 & 1.14 & 1.43 & 2.5 & 0.06 \\
017 & 2.5 & 1.94 & 1.04 & 4.0 & 47   & 059 & 3.1 & 1.60 & 1.23 & 1.0 & 0.03 & 101 & 3.1 & 1.32 & 1.44 & 2.5 & 0.03 \\
018 & 2.5 & 2.13 & 1.04 & 4.0 & 47   & 060 & 3.1 & 1.77 & 1.23 & 1.0 & 0.03 & 102 & 3.1 & 1.49 & 1.44 & 2.5 & 0.03 \\
019 & 3.1 & 0.99 & 1.03 & 1.0 & 3    & 061 & 3.1 & 0.70 & 1.19 & 2.5 & 47   & 103 & 3.1 & 0.45 & 1.34 & 4.0 & 18   \\
020 & 3.1 & 1.26 & 1.04 & 1.0 & 0.2  & 062 & 3.1 & 0.98 & 1.21 & 2.5 & 18   & 104 & 3.1 & 0.71 & 1.39 & 4.0 & 18   \\
021 & 3.1 & 1.50 & 1.04 & 1.0 & 0.06 & 063 & 3.1 & 1.22 & 1.22 & 2.5 & 18   & 105 & 3.1 & 0.95 & 1.42 & 4.0 & 18   \\
022 & 3.1 & 1.71 & 1.04 & 1.0 & 0.03 & 064 & 3.1 & 1.43 & 1.22 & 2.5 & 0.2  & 106 & 3.1 & 1.17 & 1.43 & 4.0 & 18   \\
023 & 3.1 & 1.90 & 1.04 & 1.0 & 0.03 & 065 & 3.1 & 1.62 & 1.22 & 2.5 & 0.2  & 107 & 3.1 & 1.37 & 1.44 & 4.0 & 18   \\
024 & 3.1 & 2.07 & 1.04 & 1.0 & 0.03 & 066 & 3.1 & 1.78 & 1.23 & 2.5 & 0.06 & 108 & 3.1 & 1.55 & 1.44 & 4.0 & 18   \\
025 & 3.1 & 0.99 & 1.03 & 2.5 & 47   & 067 & 3.1 & 0.72 & 1.19 & 4.0 & 47   & 109 & 2.5 & 0.28 & 1.43 & 1.0 & 3    \\
026 & 3.1 & 1.26 & 1.04 & 2.5 & 47   & 068 & 3.1 & 1.00 & 1.21 & 4.0 & 47   & 110 & 2.5 & 0.60 & 1.46 & 1.0 & 0.2  \\
027 & 3.1 & 1.51 & 1.04 & 2.5 & 47   & 069 & 3.1 & 1.25 & 1.22 & 4.0 & 47   & 111 & 2.5 & 0.79 & 1.52 & 1.0 & 0.06 \\
028 & 3.1 & 1.71 & 1.04 & 2.5 & 47   & 070 & 3.1 & 1.49 & 1.22 & 4.0 & 47   & 112 & 2.5 & 0.96 & 1.57 & 1.0 & 0.03 \\
029 & 3.1 & 1.90 & 1.04 & 2.5 & 47   & 071 & 3.1 & 1.68 & 1.23 & 4.0 & 47   & 113 & 2.5 & 1.11 & 1.61 & 1.0 & 0.03 \\
030 & 3.1 & 2.07 & 1.04 & 2.5 & 47   & 072 & 3.1 & 1.89 & 1.23 & 4.0 & 47   & 114 & 2.5 & 1.28 & 1.62 & 1.0 & 0.03 \\
031 & 3.1 & 1.01 & 1.03 & 4.0 & 47   & 073 & 2.5 & 0.53 & 1.29 & 1.0 & 3    & 115 & 2.5 & 0.33 & 1.40 & 2.5 & 3    \\
032 & 3.1 & 1.30 & 1.04 & 4.0 & 47   & 074 & 2.5 & 0.77 & 1.34 & 1.0 & 0.2  & 116 & 2.5 & 0.67 & 1.41 & 2.5 & 0.2  \\
033 & 3.1 & 1.57 & 1.04 & 4.0 & 47   & 075 & 2.5 & 0.97 & 1.39 & 1.0 & 0.06 & 117 & 2.5 & 0.87 & 1.46 & 2.5 & 0.06 \\
034 & 3.1 & 1.82 & 1.04 & 4.0 & 47   & 076 & 2.5 & 1.15 & 1.42 & 1.0 & 0.03 & 118 & 2.5 & 1.01 & 1.53 & 2.5 & 0.03 \\
035 & 3.1 & 2.06 & 1.04 & 4.0 & 47   & 077 & 2.5 & 1.33 & 1.43 & 1.0 & 0.03 & 119 & 2.5 & 1.16 & 1.57 & 2.5 & 0.03 \\
036 & 3.1 & 2.32 & 1.04 & 4.0 & 47   & 078 & 2.5 & 1.49 & 1.44 & 1.0 & 0.03 & 120 & 2.5 & 1.29 & 1.61 & 2.5 & 0.03 \\
037 & 2.5 & 0.80 & 1.14 & 1.0 & 3    & 079 & 2.5 & 0.61 & 1.24 & 2.5 & 7    & 121 & 2.5 & 0.35 & 1.39 & 4.0 & 7    \\
038 & 2.5 & 1.01 & 1.19 & 1.0 & 0.2  & 080 & 2.5 & 0.86 & 1.29 & 2.5 & 0.4  & 122 & 2.5 & 0.69 & 1.40 & 4.0 & 0.4  \\
039 & 2.5 & 1.23 & 1.21 & 1.0 & 0.06 & 081 & 2.5 & 1.04 & 1.34 & 2.5 & 0.06 & 123 & 2.5 & 0.89 & 1.45 & 4.0 & 0.2  \\
040 & 2.5 & 1.43 & 1.22 & 1.0 & 0.03 & 082 & 2.5 & 1.19 & 1.39 & 2.5 & 0.03 & 124 & 2.5 & 1.05 & 1.51 & 4.0 & 0.03 \\
041 & 2.5 & 1.62 & 1.22 & 1.0 & 0.03 & 083 & 2.5 & 1.35 & 1.42 & 2.5 & 0.03 & 125 & 2.5 & 1.18 & 1.56 & 4.0 & 0.03 \\
042 & 2.5 & 1.78 & 1.22 & 1.0 & 0.03 & 084 & 2.5 & 1.49 & 1.44 & 2.5 & 0.03 & 126 & 2.5 & 1.31 & 1.60 & 4.0 & 0.03 \\
\hline
\end{tabular}
\end{center}
\end{footnotesize}
\end{table*}
%
\end{appendix}

\end{document}